\documentclass[titlepage,aps,showpacs,prd,twocolumn,groupedaddress,superscriptaddress,floatfix]{revtex4-1}
\usepackage{graphicx}
\usepackage{dcolumn}
\usepackage{bm}
\usepackage{amsmath}
\usepackage{amsfonts}
\usepackage{comment}
\usepackage{upgreek}
\usepackage{tabularx}
\usepackage{xcolor}

\usepackage[hidelinks]{hyperref}

\begin{document}

\title{Statistical outliers in random laser emission}

\author{Federico Tommasi}\email{federico.tommasi@unifi.it} 
\author{Lorenzo Fini} 
\author{Emilio Ignesti} 
\affiliation{Dipartimento di Fisica e Astronomia, Universit\`a
di Firenze, Via Giovanni Sansone 1, I-50019 Sesto Fiorentino, Italy.} 
\author{Stefano Lepri}
\affiliation{Consiglio Nazionale delle Ricerche, Istituto dei Sistemi Complessi, Via Madonna del Piano 10, I-50019 Sesto Fiorentino, Italy.} 
\affiliation{Istituto Nazionale di Fisica Nucleare, Sezione di Firenze, 
via G. Sansone 1, I-50019 Sesto Fiorentino, Italy}
\author{Fabrizio Martelli} 
\affiliation{Dipartimento di Fisica e Astronomia, Universit\`a
di Firenze, Via Giovanni Sansone 1, I-50019 Sesto Fiorentino, Italy.} 
\author{Stefano Cavalieri} 
\affiliation{Dipartimento di Fisica e Astronomia, Universit\`a
di Firenze, Via Giovanni Sansone 1, I-50019 Sesto Fiorentino, Italy.} 

\date{\today}
\begin{abstract}
We provide theoretical and experimental evidence of statistical outliers
in random laser emission that are not accounted for by the, now established, power-law
tailed (L\'evy) distribution. Such outliers manifest themselves as single, large isolated spikes
over an otherwise smooth background.
A statistical test convincingly  shows that
their probability is larger than the one extrapolated from lower-intensity
events. To compare with experimental data, we introduced
the anomaly parameter that allows for an identification of such rare events from experimental
spectral measurements and that agrees as well with the simulations of our Monte Carlo model. A possible interpretation in terms of Black Swans or Dragon Kings, large events 
having a different generation mechanism from their peers, is discussed.
\end{abstract}

\maketitle
\section{Introduction}\label{cap:intro}
The interest in understanding and forecasting large fluctuations and extreme events 
is motivated by their tremendous impact, as it occurs in several fields, such as geology \cite{geo1}, financial crisis, social accidents \cite{economy1}, ecology \cite{ecology1} hydrology \cite{idro1} and
so on. Moreover, different optical systems have been recently studied with an approach tailored to detect extreme events and \emph{rogue waves} \cite{rw1,rw2,rw3,rw4,rw5,rw6,onorato}. Typical sources of extreme events are  complex systems, composed by a huge number of interacting entities, where underlying positive feedback mechanisms and also self-organization \cite{PhysRevLett.59.381} can act. 
In particular, a growing interest has been established in the investigation of a special class of extreme events whose appearance cannot be expected by analyzing the statistical distribution of the other observables.

Nowadays, it is well known that the statistics of many phenomena in different fields
 are described to be in good agreement with probability distributions with ``fat tail'', 
where extreme events are more likely than the normal case. 
Applications have been found in economy, where they are used to describe
 the price fluctuations, in studying the incidence and the intensity of natural
 cathastrophes \cite{terremoti1,terremoti2}, in describing paths patterns in animal foraging \cite{albatross,10.1371/journal.pcbi.1005774}, in human mobility \cite{humantravel,human} and in epidemic spread \cite{Hufnagel15124,PhysRevLett.99.188702}. Fat tail probability distribution has been studied in the context of 
anomalous diffusion \cite{fract,pl1} and demonstrated for the light propagation in 
suitably arranged disordered materials (L\'evy glasses) \cite{lf1}.  

The large-amplitude events in the ``fat tailed'' distributions can also be identified with Taleb's term \emph{Mandelbrotian Gray Swans} (GS) \cite{swan} and are thought to be caused by the same underlying mechanisms that rules the peers, despite their huge size and catastrophic impact that is beyond the realm of normal expectations. 
However, the tail-events do not tell the whole story, given also the existence of further entities called `outliers', i.e. ``anomalously'' energetic elements that appear as not described by the sample distribution. Then, such events have an extreme impact and are neither predicted nor expected by their peers. Usually, such events are known as the totally statistically intractable entities called \emph{Black Swans} (BS), after Taleb's celebrated book \cite{swan}.  Recently, Sornette \cite{dk1} introduced the intriguing different concept of \emph{Dragon King} (DK). Both DK and BW are outliers w reithspect to the distribution of the peers \cite{detDK2}, with the former that identifies a special class of outliers whose  main peculiar property is a formation mechanism distinct from the other elements of the sample \cite{dk2}. A bottleneck for the investigation 
of these phenomena is that they usually evolve by huge temporal and spatial scales,  making it difficult to achieve a long term experimental data collection.

In recent years, optical systems like random lasers \cite{rl1,rl2} proved to be versatile testbeds to 
study large fluctuations through 
well-controlled and statistically accurate experiments and numerical simulation. 

 The randomness, due to both scattering and spontaneous emission, along with the non-linearity due to the amplification mechanism play a crucial role in determining the chaotic dynamics of this kind of source. As
disordered systems, random lasers are then also an interesting platform to investigate complex out-of-equilibrium phenomena,  induced by quenched randomness like in spin glass physics \cite{spinglass1,spinglass2}. Indeed,
a remarkable replica symmetry breaking phenomenology has been demonstrated \cite{rsb1,rsb2,rsb3,rsb4}.

First experimentally realized in the 1990s \cite{rl3,rl4,rl5,rl6,sensore,Tommasi:18}, these systems are theoretically investigated and described as a huge number of interacting modes. In the diffusive regime, the main coupling mechanism among the modes is the competition for the available gain (\emph{gain coupling mechanism}) and a spiky spectrum is the result of the presence of modes that become uncoupled and promoted. The crossover from an emission characterized by smooth spectra (\emph{Gaussian regime}) to irregular, spiky ones (\emph{L\'evy regime}) can occur in a way dependent on the initial parameters, i.e.\ the disorder and the gain features \cite{lepri}. 
In the L\'evy regime, the distribution exhibits an asymptotic power-law behavior. 
  
Hence the statistics of the emission characterize the system behavior \cite{rl10,rl11,rl12,PhysRevA.90.025801,lepri,lepri2} and an extensive experimental \cite{nostro1,nostro2} and theoretical \cite{raposo2015,merrill2016} characterization of the crossover between the Gaussian and the L\'evy regimes has been reported . 
It is worth noting that the asymptotic power-law distribution of the spectral fluctuations, drawn from a system with a finite extension and stored energy, has to unavoidably undergo a cutoff \cite{PhysRevLett.73.2946,PhysRevLett.114.183903,Uppu:15} and a truncated L\'evy behavior \cite{PhysRevLett.114.183903,Uppu:15}. Recently, 
further experimental demonstration of L\'evy 
statistics has been given for random fiber lasers \cite{lima2017,lima2017extreme} and 
a lasing network of optical fibers, as a result of 
interplay  between chaotic diffusion and amplification \cite{Giacomelli}.

In such a frame, one may wonder whether statistical outliers may exist in the emission of a random laser: this 
is the object of this paper. In particular, we seek, both in the theory and in the experiment, large fluctuations whose amplitude cannot be predicted from the knowledge of smaller power-law distributed events. We will show that, albeit rare, outliers that are statistically significant are detected.
In the first part of the paper (Secs. II and III) we deal with the results of Monte Carlo simulations
while Sec. IV is devoted to the experimental results. In the last section we discuss the 
possible theoretical interpretations of the data, arguing on the distinction between Black Swans 
and Dragon Kings.

\section{Theoretical results}
In the simplest description, assuming the diffusion approximation and neglecting the gain saturation, the statistics of a random laser emission can be easily predicted.
Combining the exponential 
distribution of the random-path lengths with exponential amplification, it can be argued
that the distribution $p(I)$ of the emitted intensity (or photon number) $I$ displays a 
power-law tail \cite{lepri}:
\begin{equation}
p(I)\propto \upmu I^{-(1+\upmu)}
\label{pI}
\end{equation}
for large enough $I$. The power-law exponent  $\upmu$ being
\begin{equation}
\upmu=\frac{\ell_G}{\langle\ell\rangle}
\label{lg}
\end{equation}
where $\ell_G$ is the gain length  (the average length after which the intensity becomes amplified 
by a factor $e$) and where $\langle\ell\rangle$ is the average length of the photon path within the 
sample before being emitted through its boundaries. 

In a set of simulations with the same starting energy, one can analyze the statistics of the single events, by fitting the asymptotic behavior with a power law (with exponent $\upmu$), and also the collective behavior given by the statistical regime of the spectra (with the $\upalpha$ index of the $\upalpha$-stable fit of the spectral peaks).
Since the overall emission results from the sum of many events following the distribution (\ref{pI}),
the emission statistic is expected to follow an $\upalpha$-stable L\'evy distribution for 
$\upmu<2$ that reduces to a Gaussian in the case of finite variance for $\upmu>2$ \cite{pl1,pl2}.
Indeed, L\'evy distributions have been demonstrated to fit the experimental data and the resulting exponent
$\upalpha$ has been successfully employed to characterize large-fluctuations close to
threshold \cite{nostro1,Uppu:15,Giacomelli}.
While $\upmu$ can assume any positive value, the $\upalpha$-index attains the maximal value $\upalpha=2$ when the L\'evy distribution approaches a Gaussian. Then, when the tail of the distribution is fitted by a large value of $\upmu$ the $\upalpha$-index becomes 2. In the so-called 
L\'evy regime, $0<\upalpha<2$, we expect asymptotically $\upalpha=\upmu$ \cite{pl1,pl2}. At high energy, it is worth
noting that the large amplification and the resulting strong gain coupling mechanism between the modes, neglected in Eq.~(\ref{pI}), lead to an alteration of
the power-law trend, driving the collective behavior to a
Gaussian regime  \cite{nostro1,nostro2}.

In this section, we investigate the possibility of events that are outliers, i.e., are 
not included in the above described statistics. It is worth stressing that, in general,
an `outlier' belongs to a class of extreme events that cannot be expected by 
the distribution of the peers. To this aim, we reconsider the model previously employed in Refs.\cite{nostro1,nostro2}, whose details are given for completeness in Appendix \ref{appA}, that consists of a Monte Carlo simulation where light propagation is represented as an ensemble of random walkers propagating in a finite 2D square domain.
The main control parameter henceforth will be the initial number of excited atoms $N_0$ which
fixes the pumping level.
In this work, the number of emission spectra generated for the same starting conditions, i.~e., the same $N_0$, is 500.

Compared to the simulation carried out in the previous works \cite{nostro1,nostro2}, here the attention is focused on the statistics of the energies of \textit{individual trajectories} represented by the energy $n_i$ (or photon numbers) acquired by the $i$-th random walker at the instant in which it exits the sample ($n_i=1$ is the starting condition).  Following previous studies, we concentrate on the the survival function $S(n)=1-cdf(n)$ of each output emission where $cdf(n)$ denotes the cumulative distribution function of the random variables $n_i$.
If Eq.~(\ref{pI}) holds, we expect $S(n)$ to decay as $n^{-\upmu}$ for large $n$.

To strengthen the evidence of the existence of such outliers, we performed a statistical
test, aimed at assessing the confidence intervals of the possible deviations. 
Different statistical tests to detect outliers in a distribution have been proposed in the literature \cite{geo1,outlier,detDK1,detDK2,detDK3,negDK1}.  
Here, we used the one proposed by Janczura and Weron, based on the asymptotic properties of the empirical cumulative distribution function and the use of  the central limit theorem (see Appendix \ref{appB} and Reference \cite{detDK2} for more details). For each simulation, the values used for the power-law fit of the tail are the 0.1\% to 0.01\% largest $n_i$  over
the entire sample of $1\cdot 10^5\div1\cdot 10^6$ trajectories. We can define as a GS-like event a ``regular'' extreme value that lies in the power-law tail of the distribution. Typically, such an extreme event in a L\'evy regime causes a random spike in the emission spectrum. 	Unlike the outliers, the GSs, although they are unpredictable rare events, are expected by the distribution of the elements with lower energies. Instead, an outlier is defined as an extreme $n_i$ that lies outside the CI of 99\%.

\begin{figure}[h!]
\centerline{\scalebox{0.55}{\includegraphics{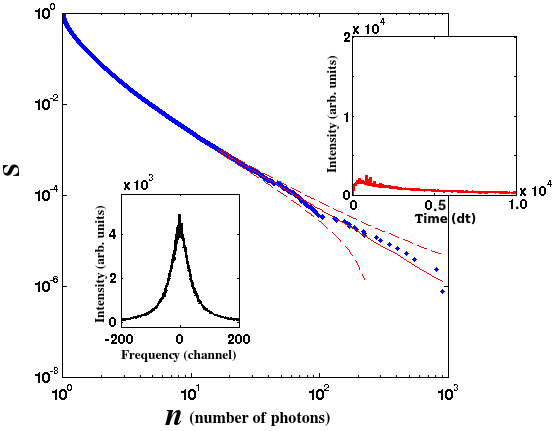}}}
\caption{(color online) Main figure: the survival function $S(n)$ for a run belonging to the set of simulations with the initial number 
$N_0=9.0\cdot 10^5$ of excited atoms and the statistical regime with $\upalpha$ = 1.56.  The 
power-law fit (0.1\% to 0.01\% largest values, solid red line) of the tail of the distribution 
yields $\upmu=1.7$. Dashed red lines represent the 99\% CI estimated as described in the 
text. No outlier outside the CI is found. In the bottom-left inset the full emission 
spectrum is shown. In the top-right inset the emission intensity for the first $10^4$ $dt$, corresponding
to the spontaneous emission lifetime $\tau$, is reported.}
\label{fig1}
\end{figure}
\begin{figure}[h!]
\centerline{\scalebox{0.55}{\includegraphics{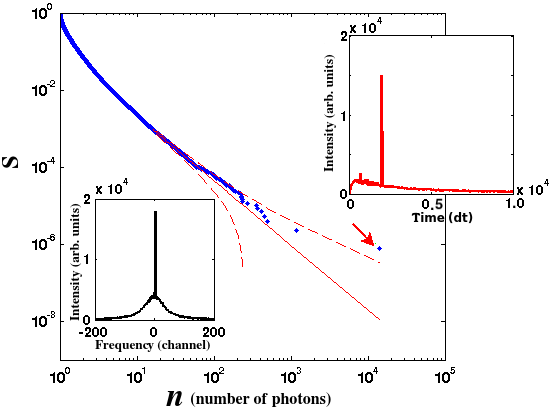}}}
\caption{(color online) Same as in the previous figure for a different simulation run having the 
same $N_0=9.0\cdot 10^5$ and $\upalpha$ = 1.56. The tail of the distribution 
yields $\upmu=1.7$. One outlier (signaled with the red arrow) clearly emerges outside the CI (dashed red lines). In the bottom-left inset, an isolated energetic spike appears on top of a smooth spectrum.  In the top-right inset an energetic single spike appears in the emission intensity, reported for the first $10^4$ $dt$, i.e.\ the spontaneous emission lifetime $\tau$. 
}
\label{fig2}
\end{figure}
\begin{figure}[h!]
\centerline{\scalebox{0.55}{\includegraphics{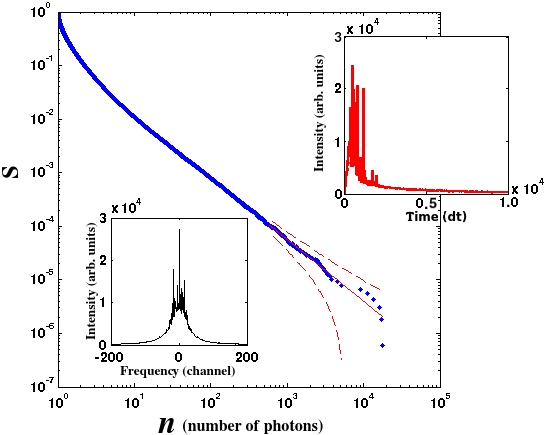}}}
\caption{(color online) Same as in the previous figures for larger initial excitation level, $N_0=1.7\cdot 10^6$ ($\upalpha$ = 1.17). 
The power-law fit (0.1\% to 0.01\% largest values, continuous line) of the tail of the distribution 
yields $\upmu=1.2$. In the bottom-left inset a spectrum with large fluctuations and random spikes is shown. However, no outlier outside the CI (dashed lines) is found and then all the extreme events are expected by the distribution of the peers. In the top-right inset the emission intensity for the first $10^4$ $dt$, i.e.\ the spontaneous emission lifetime $\tau$, is reported.
}
\label{fig3}
\end{figure} 

By the statistics of the emission spectra for runs at the same energy, it is possible to detect the statistical regime of each set of simulations. The statistical regimes of the emission as the function of energy are determined with the same 
method used in Refs.~\cite{nostro1,nostro2}, by performing the $\upalpha$-stable fit of the intensity distribution of the peak of the spectra with the same energy.

To illustrate the emergence of the outliers, we compare 
in Fig.\ \ref{fig1} and \ref{fig2} two different runs with the same starting conditions ($N_0=9.0\cdot 10^5$ excited atoms). The $\upalpha$-index for this set of simulation is 1.56, corresponding to the beginning of the L\'evy regime. At this initial energy, the typical spectra, of which the case in Fig.\ \ref{fig1} is an example, are smooth with weak fluctuations around the peak value (see the lower inset). The $S(n)$ distribution of the individual events shows that the extreme values are expected by the distribution of the weaker events. 
This situation can be compared to the case shown in Fig.\ \ref{fig2} where a single walker 
carries away an anomalously large amount of energy. We 
identify such an event as an outlier and its effect is clearly visible in the output spectrum 
as a very high and isolated peak (lower inset of Fig.\ \ref{fig2}). In the time 
domain the event is neatly detected as a single isolated burst of very large amplitude (upper inset
in  Fig.\ \ref{fig2}). The outlier shown in Fig.~\ \ref{fig2} lies distinctly outside the confidence interval (CI) of 99\%
thus confirming its special nature.  For what concerns their frequency of appearance they are very rare: in the case 
illustrated above they are detected in a few percent of the overall runs at the same energy.
 
Empirically, we found that the outliers are more likely to 
be observed at the beginning of the L\'evy region where 
$\alpha\lesssim 2$ and become harder to detect upon decreasing $\alpha$. 
Figure \ref{fig3} shows a typical case that can be obtained if the initial energy is raised to 
$N_0=1.7\cdot 10^6$ atoms, with a statistical regime that falls even more in the core of the L\'evy regime ($\upalpha=1.17$). Albeit in such a regime the fluctuations are very large, in all the 
examined cases the largest values are compatible with a (possibly) truncated power-law tail.

\section{Anomaly parameter}
 
In the experiment, the analysis of the individual events is clearly unfeasible as 
what is recorded is the overall spectral emission. To provide evidence of 
the outlier events, it is thus necessary to find an  indicator, suitable for both experimental and numerical spectra, 
capable to discriminate the character of outlier from the strongly fluctuating background. To detect the presence of an outlier, we focused on such fluctuation amplitudes that determine a `collapse' of the overall fluctuations at all other
frequencies, as happens in the numerical case of Fig.\ \ref{fig2}.
We thus introduce, for the  $i$-th spectrum, an empirical parameter $\Omega_i$ defined from the spectral shape of the emission. 
Given a sample of $N_s$ spectra with the same starting conditions, let ${I_i}(c)$ be 
the intensity of the $i$-th spectrum at a frequency identified by the frequency channel $c$. The fluctuation ${\delta_i}(c)$ is:
\begin{equation}
{\delta_i}(c)=\frac{{I_i}(c)-\langle I(c)\rangle}{\langle I(c)\rangle}\quad,
\label{delta}
\end{equation}
where 
\begin{equation}
\langle I(c)\rangle= \frac{1}{N_s}\sum_{i=1}^{N_s} I_i(c)\quad.
\label{Imedio}
\end{equation}
Then, for each spectrum, 
we evaluate the maximum 
excursion $\Omega_i'$ among the fluctuations over all $N_c$ channels as
\begin{equation}
{\Omega_i}'=|(\text{max}(\delta_{i}))^2-(\text{min}(\delta_{i}))^2|
\label{gamma_primo}
\end{equation}
To compare spectra at different energies, ${\Omega_i}'$ is normalized to the mean value over all the sample:
\begin{equation}
{\Omega_i}=\frac{{\Omega_i}'}{\langle \Omega\rangle}\quad,
\label{gamma}
\end{equation}
where $\langle \Omega\rangle$ is: 
\begin{equation}
{\langle \Omega\rangle}=\frac{1}{N_s}\sum_{i=1}^{N_s}\Omega'_i\quad.
\label{gamma_}
\end{equation}  
We call the $\Omega_i$ the \emph{anomaly parameter} since, 
by construction, relatively large values of $\Omega_i$ correspond to
spectra having individual peaks that strongly deviate from the average (like for instance 
the one in the lower inset of Fig.\ \ref{fig2}). 
It is important to  stress here that, as it is defined, the anomaly parameter
can be suitable to analyze both the experimental and the simulation data and then to compare them.

\begin{figure}[h!]
\centerline{\scalebox{0.45}{\includegraphics{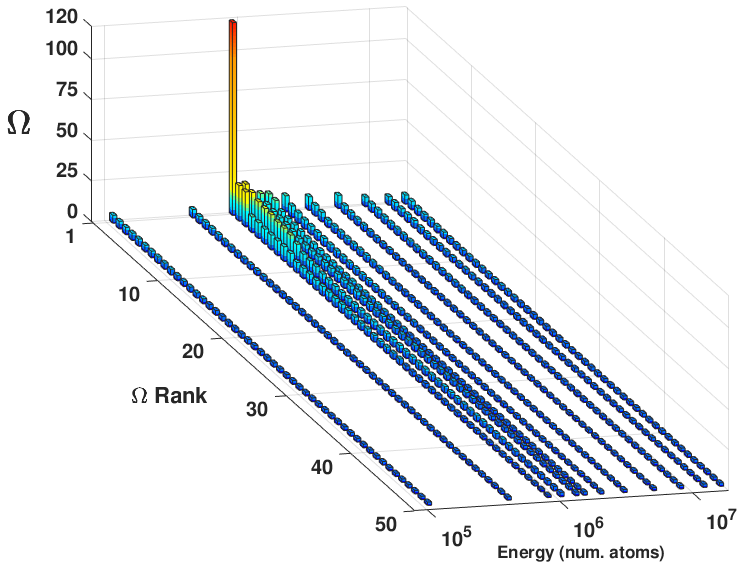}}}
\caption{(color online) Numerical data: the 50 largest values of the $\Omega_i$ parameters  
for different initial number of atoms $N_0$.
}
\label{numbar}
\end{figure}

To illustrate the effectiveness of such an indicator, let us consider first the 
simulation data. In Fig.\ \ref{numbar} the computed  $\Omega_i$ values are 
reported for different initial excitation levels $N_0$.  The $\Omega_i$ are 
ordered in decreasing order, as a function of their rank. 
The presence of the anomalous event having large values (of order $10^2$) 
is distinctly seen and they are grouped around a given initial energy. 
For instance, for the cases shown in Figs.\ \ref{fig1}, \ref{fig2} and 
\ref{fig3}, we have respectively $\Omega=1.4, 119$ and $1.5$.

\begin{figure}[h!]
\centerline{\scalebox{0.35}{\includegraphics{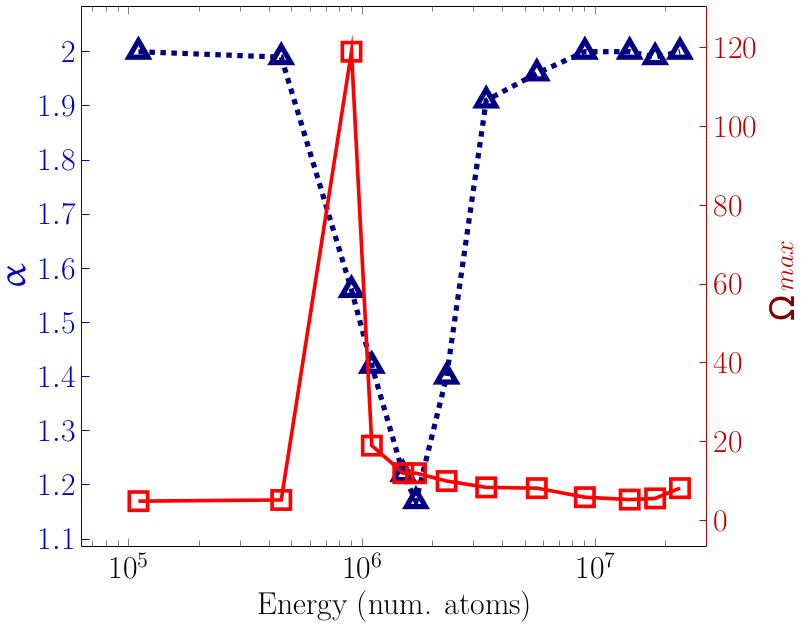}}}
\caption{(color online) Numerical data: the maximum value of $\Omega$ index reached in the whole set of spectra at the same energy (solid red line and squares, right $y$axis) and the $\upalpha$ index (dashed blue line and triangles, left $y$ axis) reported as the function of energy. The largest value of 
$\Omega_{max}$ is found in the beginning of the L\'evy regime.
}
\label{numalfabeta}
\end{figure}

In Fig.\ \ref{numalfabeta} the largest value $\Omega_{max}$, reported together with $\upalpha-$index
as a function of $N_0$, presents a peak at the beginning of the L\'evy regime, 
where $\upalpha$ starts to decrease
from 2. It is worth noting that the anomalous events are not placed in the region of minimum $\upalpha$, where the fluctuations are maximal, but are present for the values of the parameters where the spectrum is usually smooth and regular. 
 

\section{Experimental results}

We have experimentally studied the behavior of the emission spectra in the L\'evy regime. The procedure consists in collecting a large number of spectra under the same experimental conditions, in order to characterize the emission behavior, by means of the statistics of the peak intensity of the spectra. As done for the numerical simulations, the statistical regime of each set of spectra is determined. From the spectra the $\Omega$ index is also computed as described by Eq.~(\ref{gamma_}). The number of channels $c$ is fixed by the bins due to the finite resolution ($\sim 0.22$ nm) of the spectrometer. 

In the experiment, we used a sample composed by 1mM solution of Rodhamine 6G dye in ethanol with  ZnO nanoparticles to create a weakly scattering medium. The scattering mean free path $\ell_s$ at the wavelength of 632.8 nm, as measured by a method based on the Lambert-Beer law \cite{Zaccanti:03}, is 7.44 mm. The pump beam is provided by a Q-switched Nd:Yag laser, with pulse tunable in energy by a pair of polarizers. For each set of measures, an ensemble of emission spectra, produced by a pump energy that differs from the mean value by 0.5\%, were selected in order to ensure the same starting conditions. 

\begin{figure}[h!]
\centerline{\scalebox{0.6}{\includegraphics{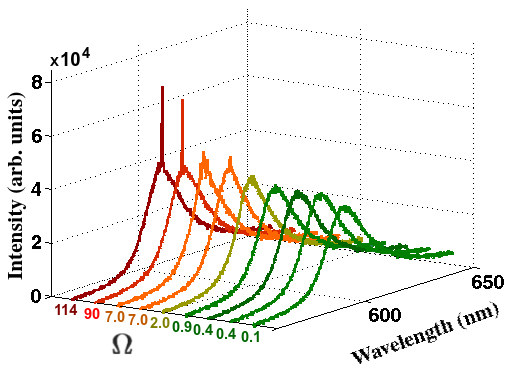}}}
\caption{(color online) Experimental spectra with pump energy of $59$ $\upmu$J ordered 
according to their respective anomaly parameters. A large value of $\Omega$ corresponds to a large spike that stands out over a smooth spectrum. The spectra with $\Omega\ll 1$ are by far the typical ones.
} 
\label{fig6}
\end{figure}

The  anomaly parameters $\Omega_i$ were calculated for each spectra: 
inspection of the data set at 59 $\upmu$J reported in Fig.\ \ref{fig6} confirms that the larger 
values of $\Omega$ correspond to energetic and isolated spikes over an otherwise smooth spectrum,
similar to what was observed in the simulations. 

\begin{figure}[h!]
\centerline{\scalebox{0.5}{\includegraphics{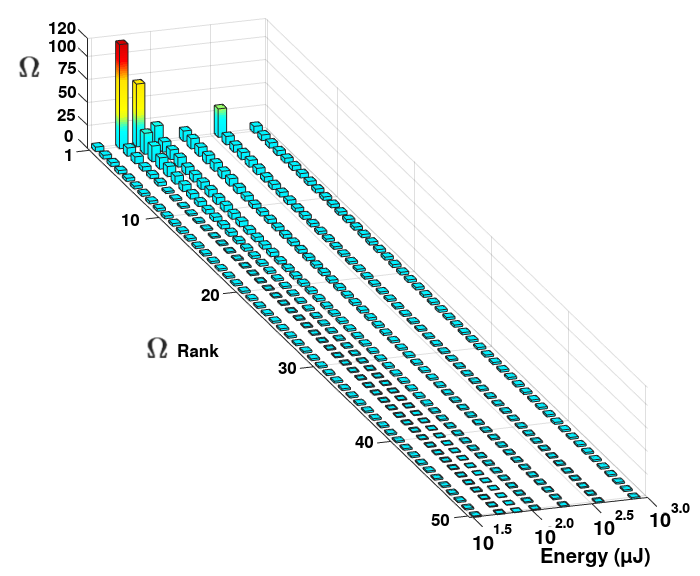}}}
\caption{(color online)
Experimental data: the 50 largest values of the $\Omega_i$ parameters  
for different pumping energies  (in $\upmu$J).
}
\label{expbar}
\end{figure}

\begin{figure}[h!]
\centerline{\scalebox{0.35}{\includegraphics{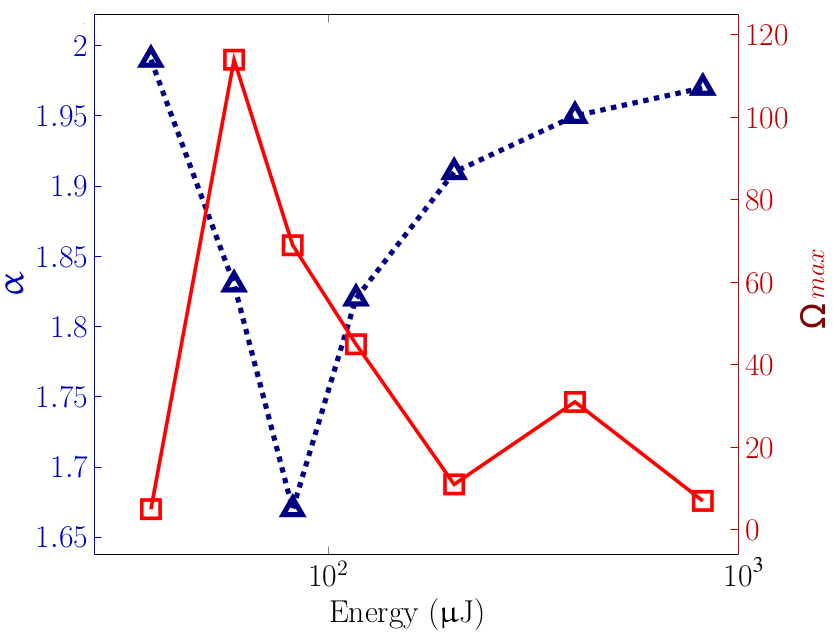}}}
\caption{(color online) Experimental data: the maximum value of $\Omega$ reached in the whole set of spectra at the same energy (solid red line and squares, right $y$ axis) and the $\upalpha$ index (dashed blue line and triangles, left $y$ axis) reported as the function of energy. The maximum value of $\Omega_{max}$ is found at the beginning of the L\'evy regime.
}
\label{alfabetaexp}
\end{figure}

Figure \ref{expbar} shows the 50 largest values of $\Omega_i$ for different pump energies, ranging from $37$ to $820$ $\upmu$J . This result appears very similar to the numerical one shown in Fig.\ \ref{numbar}. Also in this case the $\Omega$ index attains a maximum around a given pump energy.
In  Fig.\ \ref{alfabetaexp} the $\upalpha$-index and $\Omega_{max}$ are reported as a function of energy. As  in the numerical case of Fig. \ref{numalfabeta}, the largest value of $\Omega_{max}$ is 
located at the beginning of the L\'evy zone, whereas it decreases deeper inside this regime.

\begin{figure}[h!]
\centerline{\scalebox{0.35}{\includegraphics{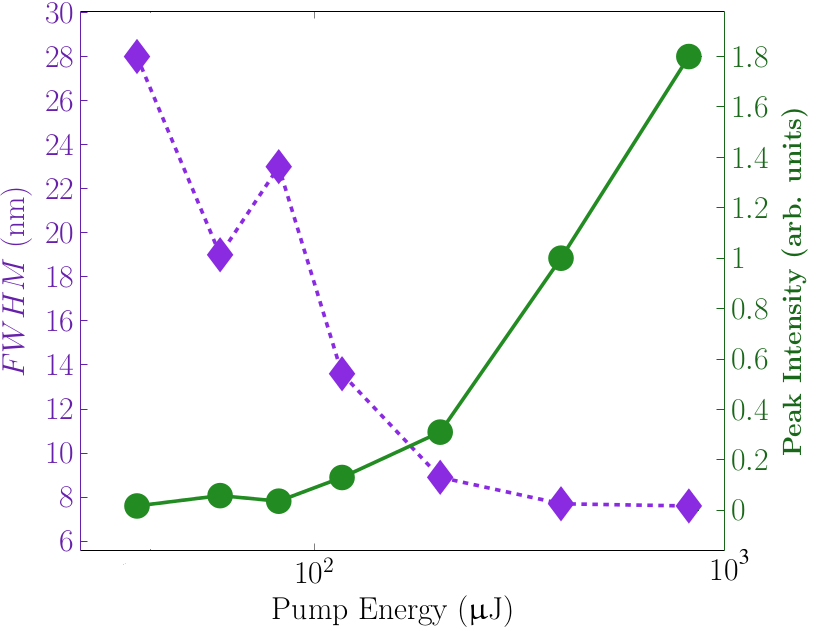}}}
\caption{(color online) Experimental data: the FWHM  (solid red line and squares, right $y$-axis) and peak value of the averaged spectrum (dashed blue line and triangles, left $y$-axis) reported as the function of energy. A threshold around $\sim 90$ $\upmu$J can be estimated by the slope changing of the peak intensity, revealing that the largest values of $\Omega$ are found in the pre-threshold energies. 
}
\label{expsoglia}
\end{figure}

By observing the change of slope of the peak intensity curve (see Fig.\ \ref{expsoglia}), the threshold energy can be estimated around $\sim90$ $\upmu$J. From this measurement, 
we conclude that the peak of the $\Omega$ index falls in a pre-threshold energy.

\section{Discussion and conclusions}  

Altogether, we have provided theoretical and experimental evidence of statistical outliers
in random laser emission that are not accounted for by the power-law
tailed distribution. Although the classification of different statistical regimes
via the $\upalpha$ index (Gaussian for $\upalpha\simeq 2$ and L\'evy for $\upalpha<2$) is undoubtedly useful \cite{nostro1,Uppu:15}, our results suggest that,
albeit in very limited parameter ranges, an even richer phenomenology may occur. 
For a quantitative analysis of experimental data, we introduced
the anomaly parameter that allows for an identification of such rare events from
spectral measurements and that agrees well with the simulations of our Monte Carlo model.
   
Having recognized the existence of statistical outliers, it remains to explain  their origin and whether indeed there is a specific generation mechanism for
them. This issue is crucial to interpret them as genuine DK as done for other systems \cite{dk1,dk2}. 
For comparison, we may refer to  the work by de Cavalcante \emph{et al.} \cite{PhysRevLett.111.198701}, that considered a  system of two coupled chaotic circuits, where extreme events arise from attractor 
bubbling pushing the trajectories away from the synchronization manifold. 
In that instance, a DK is  clearly identified by a different nonlinear 
production mechanism of large amplitude 
fluctuations, opening also the possibility to be forecasted and inhibited. 

It is thus tempting to surmise that something similar may occur also in the random laser case.
Within the theoretical model employed in this work, the possibility to identify a different cause 
of production could be only traced back to the gain competition, that is the basic nonlinear
mechanism generating gain saturation. In this respect, walkers having very long paths may 
acquire a large energy, but not larger than the one initially fed into the system 
(we recall that we are simulating an impulsive pumping).   

Being slightly above the onset of the L\'evy regime, and close 
to the random laser threshold, appears as the most advantageous condition to carry out a large amount of energy, generating a narrow peak of the spectrum, and reducing the possibility of amplification of 
the other modes. In this respect, it should be realized that a single rare event 
has a considerable impact on the whole dynamics.  

In the experiment, the evidence of outliers, unpredictable on the basis of lower intensity 
events is sound and manifests through a single peak in an otherwise smooth spectrum.
Of course,  in  the experimental case, where not all the ingredients are completely known and controlled,
the identification of a specific generation mechanism is even more difficult, 
but we may argue that a similar nonlinear effect may be a plausible cause. In the experiment, it is possible to suggest, albeit the numerical simulations show  that the outliers can have a pure statistical origin, coherent effects may lead to different production mechanisms for few events. In this scenario, besides extreme events of statistical origin, other spectral spikes due to coherent feedback mechanisms, such as randomly created optical cavities \cite{rl6,PhysRevE.61.1985} and prelocalized modes \cite{PhysRevLett.89.016802,franck}, can emerge among them in an elusive way. 
 
In conclusion, even though the identification of the outliers as genuine Dragon Kings is not a fully assessed question, and deserves further studies, our results confirm, once again, the richness of random-laser dynamics
that proves to be a relevant laboratory example for the investigation of large fluctuations 
and extreme events.

\appendix

\section{Monte Carlo model}
\label{appA}
The numerical simulations are based on the model proposed in  Refs. \cite{lepri,lepri2,nostro1,nostro2}. The dynamics of the weakly scattering random lasers can be described as an \emph{incoherent intensity feedback}, neglecting the phase information. The ``modes'' are possible random paths inside a sample and in the simulation they are described as random walkers propagating in a 2D  medium with gain, i.e., an amount of simulated excited atoms. The medium is divided in cells of side length $d\ell=vdt$, where $v$ is the velocity of a random walker and $dt$ is the temporal unit. In each cell, a number of excited atoms $N_j$ is defined.

In the simulation carried out in this work, the lattice is composed by
$150\times 150$ cells, with gain concentrated in a circular central zone with a radius of 40 cells. The stochastic behavior is then inserted by the spontaneous emission and the scattering (mean free path $\ell_s$ of 70 cells), while the non linear interactions emerge deterministically with the stimulated emission mechanism.  Once generated by a spontaneous emission event, with a probability proportional to the local population of excited atoms, 
these walkers, characterized by an individual frequency randomly drawn from a Cauchy distribution centered on the transition resonance, undergo scattering and amplification by stimulated emission. Then, the emission spectrum is generated by a spectral window consisting of $c$ frequency channels.
It is worth recalling that each simulation consists in a parallel processing of a large number of  walkers, that are thus coupled via competition for the gain.

The simulation loop consists  of three processes for each time step of duration $dt$ as follows. 
\begin{itemize}
\item[(1)] \emph{Spontaneous emission}\\
In a cell of the lattice a new random walker can be created with a probability $P_{sp}$ that is proportional to the local population of atoms $N_j$ and the spontaneous emission rate $\gamma_0$. The frequency of the walker is drawn by a Cauchy distribution of random number centered to the zero channel (the resonance), reproducing a simulated transition linewidth of $w$ arbitrary channels.
\item[(2)] \emph{Diffusion}\\
Each random walker moves to the next cell of length $d\ell$, according a trajectory that can be randomly modified by a scattering event with a probability $P_s$ per time unit.
The initial energy of the walker is one photon.
If a walker leaves the lattice, it becomes as part of the output emission. The output emission spectrum is collected by a frequency window of 1001 arbitrary channels centered on the atomic resonance (channel zero).
\item[(3)] \emph{Stimulated emission}\\
 The energy $n_i$ carried by the $i$-th random walker and the local population $N_j$ of the cell of the lattice are deterministically updated with the following rules:
\begin{eqnarray}
n_i&\to & [1+\gamma(\omega_i)dtN_j]n_i\\
N_j&\to &[1-\gamma(\omega_i)dtn_i]N_j
\label{stim2}
\end{eqnarray}

where the stimulated emission coefficient $\gamma$ depends on the random walker's frequency $\omega_i$:
\begin{equation}
\gamma(\omega_i)=\frac{\gamma_0}{1+(\omega_i/w)^2}
\label{stim3}
\end{equation}
\end{itemize} 

Given the parallel processing of a large number of walkers, the gain competition mechanism is successfully inserted. In the numerical simulations reported in this work, $\gamma_0$ is $10^{-4}$ dt$^{-1}$ (hence the spontaneous emission lifetime $\tau$ is $10^{4}$ dt) and $w$ is 50 channels of frequency. The chosen value of $P_s$ is such that the resulting $\ell_s$ is 70 $d\ell$ and each run of the simulation is long enough ($5\cdot 10^5$ $dt$) to allow the whole energy initially stored in the lattice to be extracted by the random walkers.

\section{Statistical Test}\label{appB}
The test, proposed by Janczura and Weron \cite{detDK2} and used to identify the outliers in the survival function $S(n)$, is based on the asymptotic properties of the empirical cumulative distribution function ($ecdf$) and the use of  the central limit theorem. Once the sample of $N$ elements $n_i$ is obtained, the $ecdf$ is:
\begin{equation}
ecdf(n)=\frac{1}{N}\sum_{i=1}^{N_e}\mathbb{I}_{n_i<n}
\label{test1}
\end{equation}
where $\mathbb{I}$ is the indicator function. Denoting $cdf$ the true cumulative distribution and $\mathcal{N}$ the standard normal distribution, for $N\to\infty$ the following convergence holds:
 \begin{equation}
\sqrt{N}\frac{ecdf(n)-cdf(n)}{\sqrt{cdf(n)(1-cdf(n))}}\xrightarrow{d} \mathcal{N}[0,1]
\label{test2}
\end{equation}
Let us now assume that $cdf$ follows a power law tail:
 \begin{equation}
cdf(n)\approx 1-bn^{-\upmu}\quad\text{for}\quad n\to\infty
\label{test3}
\end{equation}
where $\upmu$ is the positive power law exponent and $b$ a constant. The survival function becomes:
 \begin{equation}
S(n)=1-cdf(n)\approx bn^{-\upmu}\quad\text{for}\quad n\to\infty
\label{test4}
\end{equation}
Setting an arbitrary $k$  and indicating with $q_{k/2}$ and $q_{1-k/2}$ the $k/2$ and $1-k/2$ quantiles of $\mathcal{N}$ respectively, it can be shown that the right tail lies in the interval:
 \begin{equation}
 \begin{split}
\Bigg[&bx^{-\upmu}+\sqrt{\frac{bn^{-\upmu}(1-bn^{-\upmu})}{N}}q_{\frac{k}{2}}, \\
&bn^{-\upmu}+\sqrt{\frac{bn^{-\upmu}(1-bn^{-\upmu})}{N}}q_{1-\frac{k}{2}}\Bigg]
\end{split}
\label{test7}
\end{equation}
with probability $(1-k)$, that defines,  for each specific value of $n$, the confidence interval (CI) specifying the confidence degree for observing $ecdf(n)$ within the interval. 

Here, $k$ is fixed to 0.01 and, for each simulation, the values used for the power law fit are the 0.1\% to 0.01\% largest $n$ in complete samples of $1\cdot 10^5\div1\cdot 10^6$ elements. Then, an outlier is defined as an extreme $n_i$ that lies outside the CI of 99\%.


\begin{thebibliography}{68}%
\makeatletter
\providecommand \@ifxundefined [1]{%
 \@ifx{#1\undefined}
}%
\providecommand \@ifnum [1]{%
 \ifnum #1\expandafter \@firstoftwo
 \else \expandafter \@secondoftwo
 \fi
}%
\providecommand \@ifx [1]{%
 \ifx #1\expandafter \@firstoftwo
 \else \expandafter \@secondoftwo
 \fi
}%
\providecommand \natexlab [1]{#1}%
\providecommand \enquote  [1]{``#1''}%
\providecommand \bibnamefont  [1]{#1}%
\providecommand \bibfnamefont [1]{#1}%
\providecommand \citenamefont [1]{#1}%
\providecommand \href@noop [0]{\@secondoftwo}%
\providecommand \href [0]{\begingroup \@sanitize@url \@href}%
\providecommand \@href[1]{\@@startlink{#1}\@@href}%
\providecommand \@@href[1]{\endgroup#1\@@endlink}%
\providecommand \@sanitize@url [0]{\catcode `\\12\catcode `\$12\catcode
  `\&12\catcode `\#12\catcode `\^12\catcode `\_12\catcode `\%12\relax}%
\providecommand \@@startlink[1]{}%
\providecommand \@@endlink[0]{}%
\providecommand \url  [0]{\begingroup\@sanitize@url \@url }%
\providecommand \@url [1]{\endgroup\@href {#1}{\urlprefix }}%
\providecommand \urlprefix  [0]{URL }%
\providecommand \Eprint [0]{\href }%
\providecommand \doibase [0]{http://dx.doi.org/}%
\providecommand \selectlanguage [0]{\@gobble}%
\providecommand \bibinfo  [0]{\@secondoftwo}%
\providecommand \bibfield  [0]{\@secondoftwo}%
\providecommand \translation [1]{[#1]}%
\providecommand \BibitemOpen [0]{}%
\providecommand \bibitemStop [0]{}%
\providecommand \bibitemNoStop [0]{.\EOS\space}%
\providecommand \EOS [0]{\spacefactor3000\relax}%
\providecommand \BibitemShut  [1]{\csname bibitem#1\endcsname}%
\let\auto@bib@innerbib\@empty
\bibitem [{\citenamefont {Sachs}\ \emph
  {et~al.}(2012{\natexlab{a}})\citenamefont {Sachs}, \citenamefont {Yoder},
  \citenamefont {Turcotte}, \citenamefont {Rundle},\ and\ \citenamefont
  {Malamud}}]{geo1}%
  \BibitemOpen
  \bibfield  {author} {\bibinfo {author} {\bibfnamefont {M.~K.}\ \bibnamefont
  {Sachs}}, \bibinfo {author} {\bibfnamefont {M.~R.}\ \bibnamefont {Yoder}},
  \bibinfo {author} {\bibfnamefont {D.~L.}\ \bibnamefont {Turcotte}}, \bibinfo
  {author} {\bibfnamefont {J.~B.}\ \bibnamefont {Rundle}}, \ and\ \bibinfo
  {author} {\bibfnamefont {B.~D.}\ \bibnamefont {Malamud}},\ }\href@noop {}
  {\bibfield  {journal} {\bibinfo  {journal} {Eur. Phys. J. Special Topics}\
  }\textbf {\bibinfo {volume} {205}},\ \bibinfo {pages} {167} (\bibinfo {year}
  {2012}{\natexlab{a}})}\BibitemShut {NoStop}%
\bibitem [{\citenamefont {Schwab}\ \emph {et~al.}(2013)\citenamefont {Schwab}
  \emph {et~al.}}]{economy1}%
  \BibitemOpen
  \bibfield  {author} {\bibinfo {author} {\bibfnamefont {K.}~\bibnamefont
  {Schwab}} \emph {et~al.},\ }\href@noop {} {\emph {\bibinfo {title} {Global
  Risk}}},\ \bibinfo {edition} {8th}\ ed.\ (\bibinfo  {publisher} {World
  Economic Forum Geneva},\ \bibinfo {year} {2013})\BibitemShut {NoStop}%
\bibitem [{\citenamefont {Barnosky}\ \emph {et~al.}(2012)\citenamefont
  {Barnosky} \emph {et~al.}}]{ecology1}%
  \BibitemOpen
  \bibfield  {author} {\bibinfo {author} {\bibfnamefont {A.~D.}\ \bibnamefont
  {Barnosky}} \emph {et~al.},\ }\href@noop {} {\bibfield  {journal} {\bibinfo
  {journal} {Nature}\ }\textbf {\bibinfo {volume} {486}},\ \bibinfo {pages}
  {52} (\bibinfo {year} {2012})}\BibitemShut {NoStop}%
\bibitem [{\citenamefont {Katz}\ \emph {et~al.}(2002)\citenamefont {Katz},
  \citenamefont {Parlange},\ and\ \citenamefont {Naveau}}]{idro1}%
  \BibitemOpen
  \bibfield  {author} {\bibinfo {author} {\bibfnamefont {R.~A.}\ \bibnamefont
  {Katz}}, \bibinfo {author} {\bibfnamefont {M.~B.}\ \bibnamefont {Parlange}},
  \ and\ \bibinfo {author} {\bibfnamefont {P.}~\bibnamefont {Naveau}},\
  }\href@noop {} {\bibfield  {journal} {\bibinfo  {journal} {Adv. Water
  Resour.}\ }\textbf {\bibinfo {volume} {25}},\ \bibinfo {pages} {1287}
  (\bibinfo {year} {2002})}\BibitemShut {NoStop}%
\bibitem [{\citenamefont {Montina}\ \emph {et~al.}(2009)\citenamefont
  {Montina}, \citenamefont {Bortolozzo}, \citenamefont {Residori},\ and\
  \citenamefont {Arecchi}}]{rw1}%
  \BibitemOpen
  \bibfield  {author} {\bibinfo {author} {\bibfnamefont {A.}~\bibnamefont
  {Montina}}, \bibinfo {author} {\bibfnamefont {U.}~\bibnamefont {Bortolozzo}},
  \bibinfo {author} {\bibfnamefont {S.}~\bibnamefont {Residori}}, \ and\
  \bibinfo {author} {\bibfnamefont {F.~T.}\ \bibnamefont {Arecchi}},\ }\href
  {\doibase 10.1103/PhysRevLett.103.173901} {\bibfield  {journal} {\bibinfo
  {journal} {Phys. Rev. Lett.}\ }\textbf {\bibinfo {volume} {103}},\ \bibinfo
  {pages} {173901} (\bibinfo {year} {2009})}\BibitemShut {NoStop}%
\bibitem [{\citenamefont {Finot}\ \emph {et~al.}(2010)\citenamefont {Finot},
  \citenamefont {Hammani}, \citenamefont {Fatome}, \citenamefont {Dudley},\
  and\ \citenamefont {Millot}}]{rw2}%
  \BibitemOpen
  \bibfield  {author} {\bibinfo {author} {\bibfnamefont {C.}~\bibnamefont
  {Finot}}, \bibinfo {author} {\bibfnamefont {K.}~\bibnamefont {Hammani}},
  \bibinfo {author} {\bibfnamefont {J.}~\bibnamefont {Fatome}}, \bibinfo
  {author} {\bibfnamefont {J.~M.}\ \bibnamefont {Dudley}}, \ and\ \bibinfo
  {author} {\bibfnamefont {G.}~\bibnamefont {Millot}},\ }\href@noop {}
  {\bibfield  {journal} {\bibinfo  {journal} {IEEE Journal of Quantum
  Electronics}\ }\textbf {\bibinfo {volume} {46}},\ \bibinfo {pages} {205}
  (\bibinfo {year} {2010})}\BibitemShut {NoStop}%
\bibitem [{\citenamefont {Bonatto}\ \emph {et~al.}(2011)\citenamefont
  {Bonatto}, \citenamefont {Feyereisen}, \citenamefont {Barland}, \citenamefont
  {Giudici}, \citenamefont {Masoller}, \citenamefont {Leite},\ and\
  \citenamefont {Tredicce}}]{rw3}%
  \BibitemOpen
  \bibfield  {author} {\bibinfo {author} {\bibfnamefont {C.}~\bibnamefont
  {Bonatto}}, \bibinfo {author} {\bibfnamefont {M.}~\bibnamefont {Feyereisen}},
  \bibinfo {author} {\bibfnamefont {S.}~\bibnamefont {Barland}}, \bibinfo
  {author} {\bibfnamefont {M.}~\bibnamefont {Giudici}}, \bibinfo {author}
  {\bibfnamefont {C.}~\bibnamefont {Masoller}}, \bibinfo {author}
  {\bibfnamefont {J.~R.~R.}\ \bibnamefont {Leite}}, \ and\ \bibinfo {author}
  {\bibfnamefont {J.~R.}\ \bibnamefont {Tredicce}},\ }\href {\doibase
  10.1103/PhysRevLett.107.053901} {\bibfield  {journal} {\bibinfo  {journal}
  {Phys. Rev. Lett.}\ }\textbf {\bibinfo {volume} {107}},\ \bibinfo {pages}
  {053901} (\bibinfo {year} {2011})}\BibitemShut {NoStop}%
\bibitem [{\citenamefont {Metzger}\ \emph {et~al.}(2014)\citenamefont
  {Metzger}, \citenamefont {Fleischmann},\ and\ \citenamefont {Geisel}}]{rw4}%
  \BibitemOpen
  \bibfield  {author} {\bibinfo {author} {\bibfnamefont {J.~J.}\ \bibnamefont
  {Metzger}}, \bibinfo {author} {\bibfnamefont {R.}~\bibnamefont
  {Fleischmann}}, \ and\ \bibinfo {author} {\bibfnamefont {T.}~\bibnamefont
  {Geisel}},\ }\href {\doibase 10.1103/PhysRevLett.112.203903} {\bibfield
  {journal} {\bibinfo  {journal} {Phys. Rev. Lett.}\ }\textbf {\bibinfo
  {volume} {112}},\ \bibinfo {pages} {203903} (\bibinfo {year}
  {2014})}\BibitemShut {NoStop}%
\bibitem [{\citenamefont {Pisarchik}\ \emph {et~al.}(2011)\citenamefont
  {Pisarchik}, \citenamefont {Jaimes-Re\'ategui}, \citenamefont
  {Sevilla-Escoboza}, \citenamefont {Huerta-Cuellar},\ and\ \citenamefont
  {Taki}}]{rw5}%
  \BibitemOpen
  \bibfield  {author} {\bibinfo {author} {\bibfnamefont {A.~N.}\ \bibnamefont
  {Pisarchik}}, \bibinfo {author} {\bibfnamefont {R.}~\bibnamefont
  {Jaimes-Re\'ategui}}, \bibinfo {author} {\bibfnamefont {R.}~\bibnamefont
  {Sevilla-Escoboza}}, \bibinfo {author} {\bibfnamefont {G.}~\bibnamefont
  {Huerta-Cuellar}}, \ and\ \bibinfo {author} {\bibfnamefont {M.}~\bibnamefont
  {Taki}},\ }\href {\doibase 10.1103/PhysRevLett.107.274101} {\bibfield
  {journal} {\bibinfo  {journal} {Phys. Rev. Lett.}\ }\textbf {\bibinfo
  {volume} {107}},\ \bibinfo {pages} {274101} (\bibinfo {year}
  {2011})}\BibitemShut {NoStop}%
\bibitem [{\citenamefont {Granese}\ \emph {et~al.}(2016)\citenamefont
  {Granese}, \citenamefont {Lacapmesure}, \citenamefont {Ag\"{u}ero},
  \citenamefont {Kovalsky}, \citenamefont {Hnilo},\ and\ \citenamefont
  {Tredicce}}]{rw6}%
  \BibitemOpen
  \bibfield  {author} {\bibinfo {author} {\bibfnamefont {N.~M.}\ \bibnamefont
  {Granese}}, \bibinfo {author} {\bibfnamefont {A.}~\bibnamefont
  {Lacapmesure}}, \bibinfo {author} {\bibfnamefont {M.~B.}\ \bibnamefont
  {Ag\"{u}ero}}, \bibinfo {author} {\bibfnamefont {M.~G.}\ \bibnamefont
  {Kovalsky}}, \bibinfo {author} {\bibfnamefont {A.~A.}\ \bibnamefont {Hnilo}},
  \ and\ \bibinfo {author} {\bibfnamefont {J.~R.}\ \bibnamefont {Tredicce}},\
  }\href {\doibase 10.1364/OL.41.003010} {\bibfield  {journal} {\bibinfo
  {journal} {Opt. Lett.}\ }\textbf {\bibinfo {volume} {41}},\ \bibinfo {pages}
  {3010} (\bibinfo {year} {2016})}\BibitemShut {NoStop}%
\bibitem [{\citenamefont {Onorato}\ \emph {et~al.}(2013)\citenamefont
  {Onorato}, \citenamefont {Residori}, \citenamefont {Bortolozzo},
  \citenamefont {Montina},\ and\ \citenamefont {Arecchi}}]{onorato}%
  \BibitemOpen
  \bibfield  {author} {\bibinfo {author} {\bibfnamefont {M.}~\bibnamefont
  {Onorato}}, \bibinfo {author} {\bibfnamefont {S.}~\bibnamefont {Residori}},
  \bibinfo {author} {\bibfnamefont {U.}~\bibnamefont {Bortolozzo}}, \bibinfo
  {author} {\bibfnamefont {A.}~\bibnamefont {Montina}}, \ and\ \bibinfo
  {author} {\bibfnamefont {F.}~\bibnamefont {Arecchi}},\ }\href {\doibase
  https://doi.org/10.1016/j.physrep.2013.03.001} {\bibfield  {journal}
  {\bibinfo  {journal} {Physics Reports}\ }\textbf {\bibinfo {volume} {528}},\
  \bibinfo {pages} {47 } (\bibinfo {year} {2013})}\BibitemShut {NoStop}%
\bibitem [{\citenamefont {Bak}\ \emph {et~al.}(1987)\citenamefont {Bak},
  \citenamefont {Tang},\ and\ \citenamefont {Wiesenfeld}}]{PhysRevLett.59.381}%
  \BibitemOpen
  \bibfield  {author} {\bibinfo {author} {\bibfnamefont {P.}~\bibnamefont
  {Bak}}, \bibinfo {author} {\bibfnamefont {C.}~\bibnamefont {Tang}}, \ and\
  \bibinfo {author} {\bibfnamefont {K.}~\bibnamefont {Wiesenfeld}},\ }\href
  {\doibase 10.1103/PhysRevLett.59.381} {\bibfield  {journal} {\bibinfo
  {journal} {Phys. Rev. Lett.}\ }\textbf {\bibinfo {volume} {59}},\ \bibinfo
  {pages} {381} (\bibinfo {year} {1987})}\BibitemShut {NoStop}%
\bibitem [{\citenamefont {Sornette}\ \emph {et~al.}()\citenamefont {Sornette},
  \citenamefont {Knopoff}, \citenamefont {Kagan},\ and\ \citenamefont
  {Vanneste}}]{terremoti1}%
  \BibitemOpen
  \bibfield  {author} {\bibinfo {author} {\bibfnamefont {D.}~\bibnamefont
  {Sornette}}, \bibinfo {author} {\bibfnamefont {L.}~\bibnamefont {Knopoff}},
  \bibinfo {author} {\bibfnamefont {Y.~Y.}\ \bibnamefont {Kagan}}, \ and\
  \bibinfo {author} {\bibfnamefont {C.}~\bibnamefont {Vanneste}},\ }\href
  {\doibase 10.1029/96JB00177} {\bibfield  {journal} {\bibinfo  {journal}
  {Journal of Geophysical Research: Solid Earth}\ }\textbf {\bibinfo {volume}
  {101}},\ \bibinfo {pages} {13883} (\bibinfo {year} {1996})}\BibitemShut {NoStop}%
\bibitem [{\citenamefont {Sotolongo-Costa}\ \emph {et~al.}()\citenamefont
  {Sotolongo-Costa}, \citenamefont {Antoranz}, \citenamefont {Posadas},
  \citenamefont {Vidal},\ and\ \citenamefont {Vázquez}}]{terremoti2}%
  \BibitemOpen
  \bibfield  {author} {\bibinfo {author} {\bibfnamefont {O.}~\bibnamefont
  {Sotolongo-Costa}}, \bibinfo {author} {\bibfnamefont {J.~C.}\ \bibnamefont
  {Antoranz}}, \bibinfo {author} {\bibfnamefont {A.}~\bibnamefont {Posadas}},
  \bibinfo {author} {\bibfnamefont {F.}~\bibnamefont {Vidal}}, \ and\ \bibinfo
  {author} {\bibfnamefont {A.}~\bibnamefont {Vázquez}},\ }\href {\doibase
  10.1029/2000GL011394} {\bibfield  {journal} {\bibinfo  {journal} {Geophysical
  Research Letters}\ }\textbf {\bibinfo {volume} {27}},\ \bibinfo {pages} {1965}
  (\bibinfo {year} {2000})}\BibitemShut {NoStop}%
\bibitem [{\citenamefont {Viswanathan}\ \emph {et~al.}(1996)\citenamefont
  {Viswanathan}, \citenamefont {Afanasyev}, \citenamefont {Buldyrev},
  \citenamefont {Murphy}, \citenamefont {Prince},\ and\ \citenamefont
  {Stanley}}]{albatross}%
  \BibitemOpen
  \bibfield  {author} {\bibinfo {author} {\bibfnamefont {G.}~\bibnamefont
  {Viswanathan}}, \bibinfo {author} {\bibfnamefont {V.}~\bibnamefont
  {Afanasyev}}, \bibinfo {author} {\bibfnamefont {S.}~\bibnamefont {Buldyrev}},
  \bibinfo {author} {\bibfnamefont {E.}~\bibnamefont {Murphy}}, \bibinfo
  {author} {\bibfnamefont {P.}~\bibnamefont {Prince}}, \ and\ \bibinfo {author}
  {\bibfnamefont {H.}~\bibnamefont {Stanley}},\ }\href@noop {} {\bibfield
  {journal} {\bibinfo  {journal} {Nature}\ }\textbf {\bibinfo {volume} {381}},\
  \bibinfo {pages} {413} (\bibinfo {year} {1996})}\BibitemShut {NoStop}%
\bibitem [{\citenamefont {Wosniack}\ \emph {et~al.}(2017)\citenamefont
  {Wosniack}, \citenamefont {Santos}, \citenamefont {Raposo}, \citenamefont
  {Viswanathan},\ and\ \citenamefont {da~Luz}}]{10.1371/journal.pcbi.1005774}%
  \BibitemOpen
  \bibfield  {author} {\bibinfo {author} {\bibfnamefont {M.~E.}\ \bibnamefont
  {Wosniack}}, \bibinfo {author} {\bibfnamefont {M.~C.}\ \bibnamefont
  {Santos}}, \bibinfo {author} {\bibfnamefont {E.~P.}\ \bibnamefont {Raposo}},
  \bibinfo {author} {\bibfnamefont {G.~M.}\ \bibnamefont {Viswanathan}}, \ and\
  \bibinfo {author} {\bibfnamefont {M.~G.~E.}\ \bibnamefont {da~Luz}},\ }\href
  {\doibase 10.1371/journal.pcbi.1005774} {\bibfield  {journal} {\bibinfo
  {journal} {PLOS Computational Biology}\ }\textbf {\bibinfo {volume} {13}},\
  \bibinfo {pages} {1} (\bibinfo {year} {2017})}\BibitemShut {NoStop}%
\bibitem [{\citenamefont {Barockmann}\ \emph {et~al.}(2006)\citenamefont
  {Barockmann}, \citenamefont {Hufnagel},\ and\ \citenamefont
  {Geisel}}]{humantravel}%
  \BibitemOpen
  \bibfield  {author} {\bibinfo {author} {\bibfnamefont {D.}~\bibnamefont
  {Barockmann}}, \bibinfo {author} {\bibfnamefont {L.}~\bibnamefont
  {Hufnagel}}, \ and\ \bibinfo {author} {\bibfnamefont {T.}~\bibnamefont
  {Geisel}},\ }\href@noop {} {\bibfield  {journal} {\bibinfo  {journal}
  {Nature}\ }\textbf {\bibinfo {volume} {439}},\ \bibinfo {pages} {462}
  (\bibinfo {year} {2006})}\BibitemShut {NoStop}%
\bibitem [{\citenamefont {Rhee}\ \emph {et~al.}(2011)\citenamefont {Rhee},
  \citenamefont {Shin}, \citenamefont {Hong}, \citenamefont {Lee},
  \citenamefont {Kim},\ and\ \citenamefont {Chong}}]{human}%
  \BibitemOpen
  \bibfield  {author} {\bibinfo {author} {\bibfnamefont {I.}~\bibnamefont
  {Rhee}}, \bibinfo {author} {\bibfnamefont {M.}~\bibnamefont {Shin}}, \bibinfo
  {author} {\bibfnamefont {S.}~\bibnamefont {Hong}}, \bibinfo {author}
  {\bibfnamefont {K.}~\bibnamefont {Lee}}, \bibinfo {author} {\bibfnamefont
  {S.~J.}\ \bibnamefont {Kim}}, \ and\ \bibinfo {author} {\bibfnamefont
  {S.}~\bibnamefont {Chong}},\ }\href {\doibase 10.1109/TNET.2011.2120618}
  {\bibfield  {journal} {\bibinfo  {journal} {IEEE/ACM Transactions on
  Networking}\ }\textbf {\bibinfo {volume} {19}},\ \bibinfo {pages} {630}
  (\bibinfo {year} {2011})}\BibitemShut {NoStop}%
\bibitem [{\citenamefont {Hufnagel}\ \emph {et~al.}(2004)\citenamefont
  {Hufnagel}, \citenamefont {Brockmann},\ and\ \citenamefont
  {Geisel}}]{Hufnagel15124}%
  \BibitemOpen
  \bibfield  {author} {\bibinfo {author} {\bibfnamefont {L.}~\bibnamefont
  {Hufnagel}}, \bibinfo {author} {\bibfnamefont {D.}~\bibnamefont {Brockmann}},
  \ and\ \bibinfo {author} {\bibfnamefont {T.}~\bibnamefont {Geisel}},\ }\href
  {\doibase 10.1073/pnas.0308344101} {\bibfield  {journal} {\bibinfo  {journal}
  {Proceedings of the National Academy of Sciences}\ }\textbf {\bibinfo
  {volume} {101}},\ \bibinfo {pages} {15124} (\bibinfo {year} {2004})},\
  \Eprint
  {http://arxiv.org/abs/http://www.pnas.org/content/101/42/15124.full.pdf}
  {http://www.pnas.org/content/101/42/15124.full.pdf} \BibitemShut {NoStop}%
\bibitem [{\citenamefont {Small}\ \emph {et~al.}(2007)\citenamefont {Small},
  \citenamefont {Walker},\ and\ \citenamefont {Tse}}]{PhysRevLett.99.188702}%
  \BibitemOpen
  \bibfield  {author} {\bibinfo {author} {\bibfnamefont {M.}~\bibnamefont
  {Small}}, \bibinfo {author} {\bibfnamefont {D.~M.}\ \bibnamefont {Walker}}, \
  and\ \bibinfo {author} {\bibfnamefont {C.~K.}\ \bibnamefont {Tse}},\ }\href
  {\doibase 10.1103/PhysRevLett.99.188702} {\bibfield  {journal} {\bibinfo
  {journal} {Phys. Rev. Lett.}\ }\textbf {\bibinfo {volume} {99}},\ \bibinfo
  {pages} {188702} (\bibinfo {year} {2007})}\BibitemShut {NoStop}%
\bibitem [{\citenamefont {Mandelbrot}(1977)}]{fract}%
  \BibitemOpen
  \bibfield  {author} {\bibinfo {author} {\bibfnamefont {B.}~\bibnamefont
  {Mandelbrot}},\ }\href@noop {} {\emph {\bibinfo {title} {The Fractal Geometry
  of Nature}}}\ (\bibinfo  {publisher} {Freeman, New York (USA)},\ \bibinfo
  {year} {1977})\BibitemShut {NoStop}%
\bibitem [{\citenamefont {Bouchaud}\ and\ \citenamefont {Georges}(1990)}]{pl1}%
  \BibitemOpen
  \bibfield  {author} {\bibinfo {author} {\bibfnamefont {J.-P.}\ \bibnamefont
  {Bouchaud}}\ and\ \bibinfo {author} {\bibfnamefont {A.}~\bibnamefont
  {Georges}},\ }\href@noop {} {\bibfield  {journal} {\bibinfo  {journal} {Phys.
  Rep.}\ }\textbf {\bibinfo {volume} {195}},\ \bibinfo {pages} {127} (\bibinfo
  {year} {1990})}\BibitemShut {NoStop}%
\bibitem [{\citenamefont {Barthelemy}\ \emph {et~al.}(2008)\citenamefont
  {Barthelemy}, \citenamefont {Bertolotti},\ and\ \citenamefont
  {Wiersma}}]{lf1}%
  \BibitemOpen
  \bibfield  {author} {\bibinfo {author} {\bibfnamefont {P.}~\bibnamefont
  {Barthelemy}}, \bibinfo {author} {\bibfnamefont {J.}~\bibnamefont
  {Bertolotti}}, \ and\ \bibinfo {author} {\bibfnamefont {D.~S.}\ \bibnamefont
  {Wiersma}},\ }\href@noop {} {\bibfield  {journal} {\bibinfo  {journal}
  {Nature}\ }\textbf {\bibinfo {volume} {453}},\ \bibinfo {pages} {459}
  (\bibinfo {year} {2008})}\BibitemShut {NoStop}%
\bibitem [{\citenamefont {Taleb}(2007)}]{swan}%
  \BibitemOpen
  \bibfield  {author} {\bibinfo {author} {\bibfnamefont {N.~N.}\ \bibnamefont
  {Taleb}},\ }\href@noop {} {\emph {\bibinfo {title} {The Black Swan: The
  Impact of Highly Improbable}}}\ (\bibinfo  {publisher} {Random House, New
  York},\ \bibinfo {year} {2007})\BibitemShut {NoStop}%
\bibitem [{\citenamefont {Sornette}(2009)}]{dk1}%
  \BibitemOpen
  \bibfield  {author} {\bibinfo {author} {\bibfnamefont {D.}~\bibnamefont
  {Sornette}},\ }\href@noop {} {\bibfield  {journal} {\bibinfo  {journal}
  {ITJSE}\ }\textbf {\bibinfo {volume} {2}}(1),\ \bibinfo
  {pages} {1-18} (\bibinfo {year} {2009})}\BibitemShut {NoStop}%
\bibitem [{\citenamefont {Janczura}\ and\ \citenamefont
  {Weron}(2012)}]{detDK2}%
  \BibitemOpen
  \bibfield  {author} {\bibinfo {author} {\bibfnamefont {J.}~\bibnamefont
  {Janczura}}\ and\ \bibinfo {author} {\bibfnamefont {R.}~\bibnamefont
  {Weron}},\ }\href@noop {} {\bibfield  {journal} {\bibinfo  {journal} {Eur.
  Phys. J. Special Topic}\ }\textbf {\bibinfo {volume} {205}},\ \bibinfo
  {pages} {79} (\bibinfo {year} {2012})}\BibitemShut {NoStop}%
\bibitem [{\citenamefont {Sornette}\ and\ \citenamefont {Ouillon}(2012)}]{dk2}%
  \BibitemOpen
  \bibfield  {author} {\bibinfo {author} {\bibfnamefont {D.}~\bibnamefont
  {Sornette}}\ and\ \bibinfo {author} {\bibfnamefont {G.}~\bibnamefont
  {Ouillon}},\ }\href@noop {} {\bibfield  {journal} {\bibinfo  {journal} {Eur.
  Phys. J. Special Topic}\ }\textbf {\bibinfo {volume} {205}},\ \bibinfo
  {pages} {1} (\bibinfo {year} {2012})}\BibitemShut {NoStop}%
\bibitem [{\citenamefont {Letokhov}(1967)}]{rl1}%
  \BibitemOpen
  \bibfield  {author} {\bibinfo {author} {\bibfnamefont {V.~S.}\ \bibnamefont
  {Letokhov}},\ }\href@noop {} {\bibfield  {journal} {\bibinfo  {journal}
  {Eksp. Teor. Fiz.}\ }\textbf {\bibinfo {volume} {53}},\ \bibinfo {pages}
  {1442} (\bibinfo {year} {1967}) [Sov. Phys. JETP \textbf{26}, 835 (1968)]}\BibitemShut {NoStop}%
\bibitem [{\citenamefont {Wiersma}(2008)}]{rl2}%
  \BibitemOpen
  \bibfield  {author} {\bibinfo {author} {\bibfnamefont {D.~S.}\ \bibnamefont
  {Wiersma}},\ }\href@noop {} {\bibfield  {journal} {\bibinfo  {journal} {Nat.
  Phys.}\ }\textbf {\bibinfo {volume} {4}},\ \bibinfo {pages} {359} (\bibinfo
  {year} {2008})}\BibitemShut {NoStop}%
\bibitem [{\citenamefont {Angelani}\ \emph {et~al.}(2006)\citenamefont
  {Angelani}, \citenamefont {Conti}, \citenamefont {Ruocco},\ and\
  \citenamefont {Zamponi}}]{spinglass1}%
  \BibitemOpen
  \bibfield  {author} {\bibinfo {author} {\bibfnamefont {L.}~\bibnamefont
  {Angelani}}, \bibinfo {author} {\bibfnamefont {C.}~\bibnamefont {Conti}},
  \bibinfo {author} {\bibfnamefont {G.}~\bibnamefont {Ruocco}}, \ and\ \bibinfo
  {author} {\bibfnamefont {F.}~\bibnamefont {Zamponi}},\ }\href@noop {}
  {\bibfield  {journal} {\bibinfo  {journal} {Phys. Rev. B}\ }\textbf {\bibinfo
  {volume} {74}},\ \bibinfo {pages} {104207} (\bibinfo {year}
  {2006})}\BibitemShut {NoStop}%
\bibitem [{\citenamefont {Leuzzi}\ \emph {et~al.}(2009)\citenamefont {Leuzzi},
  \citenamefont {Conti}, \citenamefont {Folli}, \citenamefont {Angelani},\ and\
  \citenamefont {Ruocco}}]{spinglass2}%
  \BibitemOpen
  \bibfield  {author} {\bibinfo {author} {\bibfnamefont {L.}~\bibnamefont
  {Leuzzi}}, \bibinfo {author} {\bibfnamefont {C.}~\bibnamefont {Conti}},
  \bibinfo {author} {\bibfnamefont {V.}~\bibnamefont {Folli}}, \bibinfo
  {author} {\bibfnamefont {L.}~\bibnamefont {Angelani}}, \ and\ \bibinfo
  {author} {\bibfnamefont {G.}~\bibnamefont {Ruocco}},\ }\href@noop {}
  {\bibfield  {journal} {\bibinfo  {journal} {Phys. Rev. Lett.}\ }\textbf
  {\bibinfo {volume} {102}},\ \bibinfo {pages} {083901} (\bibinfo {year}
  {2009})}\BibitemShut {NoStop}%
\bibitem [{\citenamefont {Gofraniha}\ \emph {et~al.}(2015)\citenamefont
  {Gofraniha}, \citenamefont {Viola}, \citenamefont {Di~Maria}, \citenamefont
  {Barbarella}, \citenamefont {Gigli}, \citenamefont {Leuzzi},\ and\
  \citenamefont {Conti}}]{rsb1}%
  \BibitemOpen
  \bibfield  {author} {\bibinfo {author} {\bibfnamefont {N.}~\bibnamefont
  {Gofraniha}}, \bibinfo {author} {\bibfnamefont {I.}~\bibnamefont {Viola}},
  \bibinfo {author} {\bibfnamefont {F.}~\bibnamefont {Di~Maria}}, \bibinfo
  {author} {\bibfnamefont {G.}~\bibnamefont {Barbarella}}, \bibinfo {author}
  {\bibfnamefont {G.}~\bibnamefont {Gigli}}, \bibinfo {author} {\bibfnamefont
  {L.}~\bibnamefont {Leuzzi}}, \ and\ \bibinfo {author} {\bibfnamefont
  {C.}~\bibnamefont {Conti}},\ }\href@noop {} {\bibfield  {journal} {\bibinfo
  {journal} {Nat. Commun.}\ }\textbf {\bibinfo {volume} {6}},\ \bibinfo {pages}
  {6058} (\bibinfo {year} {2015})}\BibitemShut {NoStop}%
\bibitem [{\citenamefont {Antenucci}\ \emph {et~al.}(2015)\citenamefont
  {Antenucci}, \citenamefont {Conti}, \citenamefont {Ruocco},\ and\
  \citenamefont {Zamponi}}]{rsb2}%
  \BibitemOpen
  \bibfield  {author} {\bibinfo {author} {\bibfnamefont {F.}~\bibnamefont
  {Antenucci}}, \bibinfo {author} {\bibfnamefont {C.}~\bibnamefont {Conti}},
  \bibinfo {author} {\bibfnamefont {G.}~\bibnamefont {Ruocco}}, \ and\ \bibinfo
  {author} {\bibfnamefont {F.}~\bibnamefont {Zamponi}},\ }\href@noop {}
  {\bibfield  {journal} {\bibinfo  {journal} {Sci. Rep.}\ }\textbf {\bibinfo
  {volume} {5}},\ \bibinfo {pages} {16792} (\bibinfo {year}
  {2015})}\BibitemShut {NoStop}%
\bibitem [{\citenamefont {Gomes}\ \emph {et~al.}(2016)\citenamefont {Gomes},
  \citenamefont {Raposo}, \citenamefont {Moura}, \citenamefont {Fewo},
  \citenamefont {Pincheira}, \citenamefont {Jerez},\ and\ \citenamefont
  {Maia}}]{rsb3}%
  \BibitemOpen
  \bibfield  {author} {\bibinfo {author} {\bibfnamefont {A.~S.~L.}\
  \bibnamefont {Gomes}}, \bibinfo {author} {\bibfnamefont {E.~P.}\ \bibnamefont
  {Raposo}}, \bibinfo {author} {\bibfnamefont {A.~L.}\ \bibnamefont {Moura}},
  \bibinfo {author} {\bibfnamefont {S.~I.}\ \bibnamefont {Fewo}}, \bibinfo
  {author} {\bibfnamefont {P.~I.~R.}\ \bibnamefont {Pincheira}}, \bibinfo
  {author} {\bibfnamefont {V.}~\bibnamefont {Jerez}}, \ and\ \bibinfo {author}
  {\bibfnamefont {L.~J. Q. d.~A.}\ \bibnamefont {Maia}},\ }\href@noop {}
  {\bibfield  {journal} {\bibinfo  {journal} {Sci. Rep.}\ }\textbf {\bibinfo
  {volume} {5}},\ \bibinfo {pages} {27987} (\bibinfo {year}
  {2016})}\BibitemShut {NoStop}%
\bibitem [{\citenamefont {Tommasi}\ \emph {et~al.}(2016)\citenamefont
  {Tommasi}, \citenamefont {Ignesti}, \citenamefont {Lepri},\ and\
  \citenamefont {Cavalieri}}]{rsb4}%
  \BibitemOpen
  \bibfield  {author} {\bibinfo {author} {\bibfnamefont {F.}~\bibnamefont
  {Tommasi}}, \bibinfo {author} {\bibfnamefont {E.}~\bibnamefont {Ignesti}},
  \bibinfo {author} {\bibfnamefont {S.}~\bibnamefont {Lepri}}, \ and\ \bibinfo
  {author} {\bibfnamefont {S.}~\bibnamefont {Cavalieri}},\ }\href@noop {}
  {\bibfield  {journal} {\bibinfo  {journal} {Sci. Rep.}\ }\textbf {\bibinfo
  {volume} {6}},\ \bibinfo {pages} {37113} (\bibinfo {year}
  {2016})}\BibitemShut {NoStop}%
\bibitem [{\citenamefont {Lawandy}\ \emph {et~al.}(1995)\citenamefont
  {Lawandy}, \citenamefont {Balachandran}, \citenamefont {Gomes},\ and\
  \citenamefont {Suvain}}]{rl3}%
  \BibitemOpen
  \bibfield  {author} {\bibinfo {author} {\bibfnamefont {N.~M.}\ \bibnamefont
  {Lawandy}}, \bibinfo {author} {\bibfnamefont {R.~M.}\ \bibnamefont
  {Balachandran}}, \bibinfo {author} {\bibfnamefont {A.~S.~L.}\ \bibnamefont
  {Gomes}}, \ and\ \bibinfo {author} {\bibfnamefont {E.}~\bibnamefont
  {Suvain}},\ }\href@noop {} {\bibfield  {journal} {\bibinfo  {journal}
  {Nature}\ }\textbf {\bibinfo {volume} {368}},\ \bibinfo {pages} {436}
  (\bibinfo {year} {1995})}\BibitemShut {NoStop}%
\bibitem [{\citenamefont {Wiersma}\ and\ \citenamefont
  {Lagendijk}(1996)}]{rl4}%
  \BibitemOpen
  \bibfield  {author} {\bibinfo {author} {\bibfnamefont {D.~S.}\ \bibnamefont
  {Wiersma}}\ and\ \bibinfo {author} {\bibfnamefont {A.}~\bibnamefont
  {Lagendijk}},\ }\href@noop {} {\bibfield  {journal} {\bibinfo  {journal}
  {Phys. Rev. E}\ }\textbf {\bibinfo {volume} {54}},\ \bibinfo {pages} {4256}
  (\bibinfo {year} {1996})}\BibitemShut {NoStop}%
\bibitem [{\citenamefont {Noginov}\ \emph {et~al.}(1995)\citenamefont
  {Noginov}, \citenamefont {Caulfield}, \citenamefont {Noginova},\ and\
  \citenamefont {Venkateswarlu}}]{rl5}%
  \BibitemOpen
  \bibfield  {author} {\bibinfo {author} {\bibfnamefont {M.}~\bibnamefont
  {Noginov}}, \bibinfo {author} {\bibfnamefont {H.}~\bibnamefont {Caulfield}},
  \bibinfo {author} {\bibfnamefont {N.}~\bibnamefont {Noginova}}, \ and\
  \bibinfo {author} {\bibfnamefont {P.}~\bibnamefont {Venkateswarlu}},\
  }\href@noop {} {\bibfield  {journal} {\bibinfo  {journal} {Opt. Commun.}\
  }\textbf {\bibinfo {volume} {118}},\ \bibinfo {pages} {430} (\bibinfo {year}
  {1995})}\BibitemShut {NoStop}%
\bibitem [{\citenamefont {Cao}\ \emph {et~al.}(1999)\citenamefont {Cao},
  \citenamefont {Zhao}, \citenamefont {Ho}, \citenamefont {Seelig},
  \citenamefont {Wang},\ and\ \citenamefont {Chang}}]{rl6}%
  \BibitemOpen
  \bibfield  {author} {\bibinfo {author} {\bibfnamefont {H.}~\bibnamefont
  {Cao}}, \bibinfo {author} {\bibfnamefont {Y.~G.}\ \bibnamefont {Zhao}},
  \bibinfo {author} {\bibfnamefont {S.~T.}\ \bibnamefont {Ho}}, \bibinfo
  {author} {\bibfnamefont {E.~W.}\ \bibnamefont {Seelig}}, \bibinfo {author}
  {\bibfnamefont {Q.~H.}\ \bibnamefont {Wang}}, \ and\ \bibinfo {author}
  {\bibfnamefont {R.~P.~H.}\ \bibnamefont {Chang}},\ }\href@noop {} {\bibfield
  {journal} {\bibinfo  {journal} {Phys. Rev. Lett.}\ }\textbf {\bibinfo
  {volume} {82(\normalfont{11})}},\ \bibinfo {pages} {2278} (\bibinfo {year}
  {1999})}\BibitemShut {NoStop}%
\bibitem [{\citenamefont {Ignesti}\ \emph {et~al.}(2016)\citenamefont
  {Ignesti}, \citenamefont {Tommasi}, \citenamefont {Fini}, \citenamefont
  {Martelli}, \citenamefont {Azzali},\ and\ \citenamefont
  {Cavalieri}}]{sensore}%
  \BibitemOpen
  \bibfield  {author} {\bibinfo {author} {\bibfnamefont {E.}~\bibnamefont
  {Ignesti}}, \bibinfo {author} {\bibfnamefont {F.}~\bibnamefont {Tommasi}},
  \bibinfo {author} {\bibfnamefont {L.}~\bibnamefont {Fini}}, \bibinfo {author}
  {\bibfnamefont {F.}~\bibnamefont {Martelli}}, \bibinfo {author}
  {\bibfnamefont {N.}~\bibnamefont {Azzali}}, \ and\ \bibinfo {author}
  {\bibfnamefont {S.}~\bibnamefont {Cavalieri}},\ }\href@noop {} {\bibfield
  {journal} {\bibinfo  {journal} {Sci. Rep.}\ }\textbf {\bibinfo {volume}
  {6}},\ \bibinfo {pages} {35225} (\bibinfo {year} {2016})}\BibitemShut
  {NoStop}%
\bibitem [{\citenamefont {Tommasi}\ \emph {et~al.}(2018)\citenamefont
  {Tommasi}, \citenamefont {Ignesti}, \citenamefont {Fini}, \citenamefont
  {Martelli},\ and\ \citenamefont {Cavalieri}}]{Tommasi:18}%
  \BibitemOpen
  \bibfield  {author} {\bibinfo {author} {\bibfnamefont {F.}~\bibnamefont
  {Tommasi}}, \bibinfo {author} {\bibfnamefont {E.}~\bibnamefont {Ignesti}},
  \bibinfo {author} {\bibfnamefont {L.}~\bibnamefont {Fini}}, \bibinfo {author}
  {\bibfnamefont {F.}~\bibnamefont {Martelli}}, \ and\ \bibinfo {author}
  {\bibfnamefont {S.}~\bibnamefont {Cavalieri}},\ }\href {\doibase
  10.1364/OE.26.027615} {\bibfield  {journal} {\bibinfo  {journal} {Opt.
  Express}\ }\textbf {\bibinfo {volume} {26}},\ \bibinfo {pages} {27615}
  (\bibinfo {year} {2018})}\BibitemShut {NoStop}%
\bibitem [{\citenamefont {Lepri}\ \emph {et~al.}(2007)\citenamefont {Lepri},
  \citenamefont {Cavalieri}, \citenamefont {Oppo},\ and\ \citenamefont
  {Wiersma}}]{lepri}%
  \BibitemOpen
  \bibfield  {author} {\bibinfo {author} {\bibfnamefont {S.}~\bibnamefont
  {Lepri}}, \bibinfo {author} {\bibfnamefont {S.}~\bibnamefont {Cavalieri}},
  \bibinfo {author} {\bibfnamefont {G.~L.}\ \bibnamefont {Oppo}}, \ and\
  \bibinfo {author} {\bibfnamefont {D.~S.}\ \bibnamefont {Wiersma}},\
  }\href@noop {} {\bibfield  {journal} {\bibinfo  {journal} {Phys. Rev. A}\
  }\textbf {\bibinfo {volume} {75}},\ \bibinfo {pages} {063820} (\bibinfo
  {year} {2007})}\BibitemShut {NoStop}%
\bibitem [{\citenamefont {Sharma}\ \emph {et~al.}(2006)\citenamefont {Sharma},
  \citenamefont {Ramachandran},\ and\ \citenamefont {Kumar}}]{rl10}%
  \BibitemOpen
  \bibfield  {author} {\bibinfo {author} {\bibfnamefont {D.}~\bibnamefont
  {Sharma}}, \bibinfo {author} {\bibfnamefont {H.}~\bibnamefont
  {Ramachandran}}, \ and\ \bibinfo {author} {\bibfnamefont {N.}~\bibnamefont
  {Kumar}},\ }\href@noop {} {\bibfield  {journal} {\bibinfo  {journal} {Fluc.
  Noise Lett.}\ }\textbf {\bibinfo {volume} {6}},\ \bibinfo {pages} {L95}
  (\bibinfo {year} {2006})}\BibitemShut {NoStop}%
\bibitem [{\citenamefont {Uppu}\ \emph {et~al.}(2012)\citenamefont {Uppu},
  \citenamefont {Tiwari},\ and\ \citenamefont {Mujumdar}}]{rl11}%
  \BibitemOpen
  \bibfield  {author} {\bibinfo {author} {\bibfnamefont {R.}~\bibnamefont
  {Uppu}}, \bibinfo {author} {\bibfnamefont {A.~K.}\ \bibnamefont {Tiwari}}, \
  and\ \bibinfo {author} {\bibfnamefont {S.}~\bibnamefont {Mujumdar}},\
  }\href@noop {} {\bibfield  {journal} {\bibinfo  {journal} {Opt. Lett.}\
  }\textbf {\bibinfo {volume} {37(\normalfont{4})}},\ \bibinfo {pages} {662}
  (\bibinfo {year} {2012})}\BibitemShut {NoStop}%
\bibitem [{\citenamefont {Uppu}\ and\ \citenamefont {Mujumdar}(2013)}]{rl12}%
  \BibitemOpen
  \bibfield  {author} {\bibinfo {author} {\bibfnamefont {R.}~\bibnamefont
  {Uppu}}\ and\ \bibinfo {author} {\bibfnamefont {S.}~\bibnamefont
  {Mujumdar}},\ }\href@noop {} {\bibfield  {journal} {\bibinfo  {journal}
  {Phys. Rev. A}\ }\textbf {\bibinfo {volume} {87}},\ \bibinfo {pages} {013822}
  (\bibinfo {year} {2013})}\BibitemShut {NoStop}%
\bibitem [{\citenamefont {Uppu}\ and\ \citenamefont
  {Mujumdar}(2014)}]{PhysRevA.90.025801}%
  \BibitemOpen
  \bibfield  {author} {\bibinfo {author} {\bibfnamefont {R.}~\bibnamefont
  {Uppu}}\ and\ \bibinfo {author} {\bibfnamefont {S.}~\bibnamefont
  {Mujumdar}},\ }\href {\doibase 10.1103/PhysRevA.90.025801} {\bibfield
  {journal} {\bibinfo  {journal} {Phys. Rev. A}\ }\textbf {\bibinfo {volume}
  {90}},\ \bibinfo {pages} {025801} (\bibinfo {year} {2014})}\BibitemShut
  {NoStop}%
\bibitem [{\citenamefont {Lepri}(2013)}]{lepri2}%
  \BibitemOpen
  \bibfield  {author} {\bibinfo {author} {\bibfnamefont {S.}~\bibnamefont
  {Lepri}},\ }\href@noop {} {\bibfield  {journal} {\bibinfo  {journal} {Phys.
  Rev. Lett.}\ }\textbf {\bibinfo {volume} {110}},\ \bibinfo {pages} {230603}
  (\bibinfo {year} {2013})}\BibitemShut {NoStop}%
\bibitem [{\citenamefont {Ignesti}\ \emph {et~al.}(2013)\citenamefont
  {Ignesti}, \citenamefont {Tommasi}, \citenamefont {Fini}, \citenamefont
  {Lepri}, \citenamefont {Radhalakshmi}, \citenamefont {Wiersma},\ and\
  \citenamefont {Cavalieri}}]{nostro1}%
  \BibitemOpen
  \bibfield  {author} {\bibinfo {author} {\bibfnamefont {E.}~\bibnamefont
  {Ignesti}}, \bibinfo {author} {\bibfnamefont {F.}~\bibnamefont {Tommasi}},
  \bibinfo {author} {\bibfnamefont {L.}~\bibnamefont {Fini}}, \bibinfo {author}
  {\bibfnamefont {S.}~\bibnamefont {Lepri}}, \bibinfo {author} {\bibfnamefont
  {V.}~\bibnamefont {Radhalakshmi}}, \bibinfo {author} {\bibfnamefont {D.~S.}\
  \bibnamefont {Wiersma}}, \ and\ \bibinfo {author} {\bibfnamefont
  {S.}~\bibnamefont {Cavalieri}},\ }\href@noop {} {\bibfield  {journal}
  {\bibinfo  {journal} {Phys. Rev. A}\ }\textbf {\bibinfo {volume} {88}},\
  \bibinfo {pages} {033820} (\bibinfo {year} {2013})}\BibitemShut {NoStop}%
\bibitem [{\citenamefont {Tommasi}\ \emph {et~al.}(2015)\citenamefont
  {Tommasi}, \citenamefont {Ignesti}, \citenamefont {Fini},\ and\ \citenamefont
  {Cavalieri}}]{nostro2}%
  \BibitemOpen
  \bibfield  {author} {\bibinfo {author} {\bibfnamefont {F.}~\bibnamefont
  {Tommasi}}, \bibinfo {author} {\bibfnamefont {E.}~\bibnamefont {Ignesti}},
  \bibinfo {author} {\bibfnamefont {L.}~\bibnamefont {Fini}}, \ and\ \bibinfo
  {author} {\bibfnamefont {S.}~\bibnamefont {Cavalieri}},\ }\href@noop {}
  {\bibfield  {journal} {\bibinfo  {journal} {Phys. Rev. A}\ }\textbf {\bibinfo
  {volume} {91}},\ \bibinfo {pages} {033820} (\bibinfo {year}
  {2015})}\BibitemShut {NoStop}%
\bibitem [{\citenamefont {Raposo}\ and\ \citenamefont
  {Gomes}(2015)}]{raposo2015}%
  \BibitemOpen
  \bibfield  {author} {\bibinfo {author} {\bibfnamefont {E.~P.}\ \bibnamefont
  {Raposo}}\ and\ \bibinfo {author} {\bibfnamefont {A.~S.~L.}\ \bibnamefont
  {Gomes}},\ }\href {\doibase 10.1103/PhysRevA.91.043827} {\bibfield  {journal}
  {\bibinfo  {journal} {Phys. Rev. A}\ }\textbf {\bibinfo {volume} {91}},\
  \bibinfo {pages} {043827} (\bibinfo {year} {2015})}\BibitemShut {NoStop}%
\bibitem [{\citenamefont {Merrill}\ \emph {et~al.}(2016)\citenamefont
  {Merrill}, \citenamefont {Cao},\ and\ \citenamefont
  {Dufresne}}]{merrill2016}%
  \BibitemOpen
  \bibfield  {author} {\bibinfo {author} {\bibfnamefont {J.~W.}\ \bibnamefont
  {Merrill}}, \bibinfo {author} {\bibfnamefont {H.}~\bibnamefont {Cao}}, \ and\
  \bibinfo {author} {\bibfnamefont {E.~R.}\ \bibnamefont {Dufresne}},\ }\href
  {\doibase 10.1103/PhysRevA.93.021801} {\bibfield  {journal} {\bibinfo
  {journal} {Phys. Rev. A}\ }\textbf {\bibinfo {volume} {93}},\ \bibinfo
  {pages} {021801} (\bibinfo {year} {2016})}\BibitemShut {NoStop}%
\bibitem [{\citenamefont {Mantegna}\ and\ \citenamefont
  {Stanley}(1994)}]{PhysRevLett.73.2946}%
  \BibitemOpen
  \bibfield  {author} {\bibinfo {author} {\bibfnamefont {R.~N.}\ \bibnamefont
  {Mantegna}}\ and\ \bibinfo {author} {\bibfnamefont {H.~E.}\ \bibnamefont
  {Stanley}},\ }\href {\doibase 10.1103/PhysRevLett.73.2946} {\bibfield
  {journal} {\bibinfo  {journal} {Phys. Rev. Lett.}\ }\textbf {\bibinfo
  {volume} {73}},\ \bibinfo {pages} {2946} (\bibinfo {year}
  {1994})}\BibitemShut {NoStop}%
\bibitem [{\citenamefont {Uppu}\ and\ \citenamefont
  {Mujumdar}(2015{\natexlab{a}})}]{PhysRevLett.114.183903}%
  \BibitemOpen
  \bibfield  {author} {\bibinfo {author} {\bibfnamefont {R.}~\bibnamefont
  {Uppu}}\ and\ \bibinfo {author} {\bibfnamefont {S.}~\bibnamefont
  {Mujumdar}},\ }\href {\doibase 10.1103/PhysRevLett.114.183903} {\bibfield
  {journal} {\bibinfo  {journal} {Phys. Rev. Lett.}\ }\textbf {\bibinfo
  {volume} {114}},\ \bibinfo {pages} {183903} (\bibinfo {year}
  {2015}{\natexlab{a}})}\BibitemShut {NoStop}%
\bibitem [{\citenamefont {Uppu}\ and\ \citenamefont
  {Mujumdar}(2015{\natexlab{b}})}]{Uppu:15}%
  \BibitemOpen
  \bibfield  {author} {\bibinfo {author} {\bibfnamefont {R.}~\bibnamefont
  {Uppu}}\ and\ \bibinfo {author} {\bibfnamefont {S.}~\bibnamefont
  {Mujumdar}},\ }\href {\doibase 10.1364/OL.40.005046} {\bibfield  {journal}
  {\bibinfo  {journal} {Opt. Lett.}\ }\textbf {\bibinfo {volume} {40}},\
  \bibinfo {pages} {5046} (\bibinfo {year} {2015}{\natexlab{b}})}\BibitemShut
  {NoStop}%
\bibitem [{\citenamefont {Lima}\ \emph
  {et~al.}(2017{\natexlab{a}})\citenamefont {Lima}, \citenamefont {Gomes},
  \citenamefont {Pincheira}, \citenamefont {Moura}, \citenamefont {Gagn{\'e}},
  \citenamefont {Raposo}, \citenamefont {de~Ara{\'u}jo},\ and\ \citenamefont
  {Kashyap}}]{lima2017}%
  \BibitemOpen
  \bibfield  {author} {\bibinfo {author} {\bibfnamefont {B.~C.}\ \bibnamefont
  {Lima}}, \bibinfo {author} {\bibfnamefont {A.~S.}\ \bibnamefont {Gomes}},
  \bibinfo {author} {\bibfnamefont {P.~I.}\ \bibnamefont {Pincheira}}, \bibinfo
  {author} {\bibfnamefont {A.~L.}\ \bibnamefont {Moura}}, \bibinfo {author}
  {\bibfnamefont {M.}~\bibnamefont {Gagn{\'e}}}, \bibinfo {author}
  {\bibfnamefont {E.~P.}\ \bibnamefont {Raposo}}, \bibinfo {author}
  {\bibfnamefont {C.~B.}\ \bibnamefont {de~Ara{\'u}jo}}, \ and\ \bibinfo
  {author} {\bibfnamefont {R.}~\bibnamefont {Kashyap}},\ }\href@noop {}
  {\bibfield  {journal} {\bibinfo  {journal} {JOSA B}\ }\textbf {\bibinfo
  {volume} {34}},\ \bibinfo {pages} {293} (\bibinfo {year}
  {2017}{\natexlab{a}})}\BibitemShut {NoStop}%
\bibitem [{\citenamefont {Lima}\ \emph
  {et~al.}(2017{\natexlab{b}})\citenamefont {Lima}, \citenamefont {Pincheira},
  \citenamefont {Raposo}, \citenamefont {Menezes}, \citenamefont {de~Ara\'ujo},
  \citenamefont {Gomes},\ and\ \citenamefont {Kashyap}}]{lima2017extreme}%
  \BibitemOpen
  \bibfield  {author} {\bibinfo {author} {\bibfnamefont {B.~C.}\ \bibnamefont
  {Lima}}, \bibinfo {author} {\bibfnamefont {P.~I.~R.}\ \bibnamefont
  {Pincheira}}, \bibinfo {author} {\bibfnamefont {E.~P.}\ \bibnamefont
  {Raposo}}, \bibinfo {author} {\bibfnamefont {L.~d.~S.}\ \bibnamefont
  {Menezes}}, \bibinfo {author} {\bibfnamefont {C.~B.}\ \bibnamefont
  {de~Ara\'ujo}}, \bibinfo {author} {\bibfnamefont {A.~S.~L.}\ \bibnamefont
  {Gomes}}, \ and\ \bibinfo {author} {\bibfnamefont {R.}~\bibnamefont
  {Kashyap}},\ }\href {\doibase 10.1103/PhysRevA.96.013834} {\bibfield
  {journal} {\bibinfo  {journal} {Phys. Rev. A}\ }\textbf {\bibinfo {volume}
  {96}},\ \bibinfo {pages} {013834} (\bibinfo {year}
  {2017}{\natexlab{b}})}\BibitemShut {NoStop}%
\bibitem [{\citenamefont {Lepri}\ \emph {et~al.}(2017)\citenamefont {Lepri},
  \citenamefont {Trono},\ and\ \citenamefont {Giacomelli}}]{Giacomelli}%
  \BibitemOpen
  \bibfield  {author} {\bibinfo {author} {\bibfnamefont {S.}~\bibnamefont
  {Lepri}}, \bibinfo {author} {\bibfnamefont {C.}~\bibnamefont {Trono}}, \ and\
  \bibinfo {author} {\bibfnamefont {G.}~\bibnamefont {Giacomelli}},\ }\href
  {\doibase 10.1103/PhysRevLett.118.123901} {\bibfield  {journal} {\bibinfo
  {journal} {Phys. Rev. Lett.}\ }\textbf {\bibinfo {volume} {118}},\ \bibinfo
  {pages} {123901} (\bibinfo {year} {2017})}\BibitemShut {NoStop}%
\bibitem [{\citenamefont {Samorodnitsky}\ and\ \citenamefont
  {Taqqu}(1994)}]{pl2}%
  \BibitemOpen
  \bibfield  {author} {\bibinfo {author} {\bibfnamefont {G.}~\bibnamefont
  {Samorodnitsky}}\ and\ \bibinfo {author} {\bibfnamefont {M.}~\bibnamefont
  {Taqqu}},\ }\href@noop {} {\emph {\bibinfo {title} {Stable Non-Gaussian
  Random Processes: Stochastic Models with Infinite Variance}}}\ (\bibinfo
  {publisher} {Chapman and Hall/CNR, New York},\ \bibinfo {year}
  {1994})\BibitemShut {NoStop}%
\bibitem [{\citenamefont {Grubbs}(1969)}]{outlier}%
  \BibitemOpen
  \bibfield  {author} {\bibinfo {author} {\bibfnamefont {F.~E.}\ \bibnamefont
  {Grubbs}},\ }\href {\doibase 10.1080/00401706.1969.10490657} {\bibfield
  {journal} {\bibinfo  {journal} {Technometrics}\ }\textbf {\bibinfo {volume}
  {11}},\ \bibinfo {pages} {1} (\bibinfo {year} {1969})}\BibitemShut {NoStop}%
\bibitem [{\citenamefont {Pisarenko}\ and\ \citenamefont
  {Sornette}(2012)}]{detDK1}%
  \BibitemOpen
  \bibfield  {author} {\bibinfo {author} {\bibfnamefont {V.~F.}\ \bibnamefont
  {Pisarenko}}\ and\ \bibinfo {author} {\bibfnamefont {D.}~\bibnamefont
  {Sornette}},\ }\href@noop {} {\bibfield  {journal} {\bibinfo  {journal} {Eur.
  Phys. J. Special Topic}\ }\textbf {\bibinfo {volume} {205}},\ \bibinfo
  {pages} {95} (\bibinfo {year} {2012})}\BibitemShut {NoStop}%
\bibitem [{\citenamefont {Ancey}(2012)}]{detDK3}%
  \BibitemOpen
  \bibfield  {author} {\bibinfo {author} {\bibfnamefont {C.}~\bibnamefont
  {Ancey}},\ }\href@noop {} {\bibfield  {journal} {\bibinfo  {journal} {Eur.
  Phys. J. Special Topic}\ }\textbf {\bibinfo {volume} {205}},\ \bibinfo
  {pages} {117} (\bibinfo {year} {2012})}\BibitemShut {NoStop}%
\bibitem [{\citenamefont {Riva}\ \emph {et~al.}(2013)\citenamefont {Riva},
  \citenamefont {Neuman},\ and\ \citenamefont {Guadagnini}}]{negDK1}%
  \BibitemOpen
  \bibfield  {author} {\bibinfo {author} {\bibfnamefont {M.}~\bibnamefont
  {Riva}}, \bibinfo {author} {\bibfnamefont {S.~P.}\ \bibnamefont {Neuman}}, \
  and\ \bibinfo {author} {\bibfnamefont {A.}~\bibnamefont {Guadagnini}},\
  }\href@noop {} {\bibfield  {journal} {\bibinfo  {journal} {Nonlin. Processes
  Geophys}\ }\textbf {\bibinfo {volume} {20}},\ \bibinfo {pages} {549}
  (\bibinfo {year} {2013})}\BibitemShut {NoStop}%
\bibitem [{\citenamefont {Zaccanti}\ \emph {et~al.}(2003)\citenamefont
  {Zaccanti}, \citenamefont {{Del Bianco}},\ and\ \citenamefont
  {Martelli}}]{Zaccanti:03}%
  \BibitemOpen
  \bibfield  {author} {\bibinfo {author} {\bibfnamefont {G.}~\bibnamefont
  {Zaccanti}}, \bibinfo {author} {\bibfnamefont {S.}~\bibnamefont {{Del
  Bianco}}}, \ and\ \bibinfo {author} {\bibfnamefont {F.}~\bibnamefont
  {Martelli}},\ }\href {\doibase 10.1364/AO.42.004023} {\bibfield  {journal}
  {\bibinfo  {journal} {Appl. Opt.}\ }\textbf {\bibinfo {volume} {42}},\
  \bibinfo {pages} {4023} (\bibinfo {year} {2003})}\BibitemShut {NoStop}%
\bibitem [{\citenamefont {de~S.~Cavalcante}\ \emph {et~al.}(2013)\citenamefont
  {de~S.~Cavalcante}, \citenamefont {Ori\'a}, \citenamefont {Sornette},
  \citenamefont {Ott},\ and\ \citenamefont
  {Gauthier}}]{PhysRevLett.111.198701}%
  \BibitemOpen
  \bibfield  {author} {\bibinfo {author} {\bibfnamefont {H.}~\bibnamefont
  {de~S.~Cavalcante}}, \bibinfo {author} {\bibfnamefont {M.}~\bibnamefont
  {Ori\'a}}, \bibinfo {author} {\bibfnamefont {D.}~\bibnamefont {Sornette}},
  \bibinfo {author} {\bibfnamefont {E.}~\bibnamefont {Ott}}, \ and\ \bibinfo
  {author} {\bibfnamefont {D.~J.}\ \bibnamefont {Gauthier}},\ }\href {\doibase
  10.1103/PhysRevLett.111.198701} {\bibfield  {journal} {\bibinfo  {journal}
  {Phys. Rev. Lett.}\ }\textbf {\bibinfo {volume} {111}},\ \bibinfo {pages}
  {198701} (\bibinfo {year} {2013})}\BibitemShut {NoStop}%
\bibitem [{\citenamefont {Cao}\ \emph {et~al.}(2000)\citenamefont {Cao},
  \citenamefont {Xu}, \citenamefont {Chang},\ and\ \citenamefont
  {Ho}}]{PhysRevE.61.1985}%
  \BibitemOpen
  \bibfield  {author} {\bibinfo {author} {\bibfnamefont {H.}~\bibnamefont
  {Cao}}, \bibinfo {author} {\bibfnamefont {J.~Y.}\ \bibnamefont {Xu}},
  \bibinfo {author} {\bibfnamefont {S.-H.}\ \bibnamefont {Chang}}, \ and\
  \bibinfo {author} {\bibfnamefont {S.~T.}\ \bibnamefont {Ho}},\ }\href
  {\doibase 10.1103/PhysRevE.61.1985} {\bibfield  {journal} {\bibinfo
  {journal} {Phys. Rev. E}\ }\textbf {\bibinfo {volume} {61}},\ \bibinfo
  {pages} {1985} (\bibinfo {year} {2000})}\BibitemShut {NoStop}%
\bibitem [{\citenamefont {Apalkov}\ \emph {et~al.}(2002)\citenamefont
  {Apalkov}, \citenamefont {Raikh},\ and\ \citenamefont
  {Shapiro}}]{PhysRevLett.89.016802}%
  \BibitemOpen
  \bibfield  {author} {\bibinfo {author} {\bibfnamefont {V.~M.}\ \bibnamefont
  {Apalkov}}, \bibinfo {author} {\bibfnamefont {M.~E.}\ \bibnamefont {Raikh}},
  \ and\ \bibinfo {author} {\bibfnamefont {B.}~\bibnamefont {Shapiro}},\ }\href
  {\doibase 10.1103/PhysRevLett.89.016802} {\bibfield  {journal} {\bibinfo
  {journal} {Phys. Rev. Lett.}\ }\textbf {\bibinfo {volume} {89}},\ \bibinfo
  {pages} {016802} (\bibinfo {year} {2002})}\BibitemShut {NoStop}%
\bibitem [{\citenamefont {Franck}\ \emph {et~al.}(2009)\citenamefont {Franck},
  \citenamefont {Lubatsch},\ and\ \citenamefont {Kroha}}]{franck}%
  \BibitemOpen
  \bibfield  {author} {\bibinfo {author} {\bibfnamefont {R.}~\bibnamefont
  {Franck}}, \bibinfo {author} {\bibfnamefont {A.}~\bibnamefont {Lubatsch}}, \
  and\ \bibinfo {author} {\bibfnamefont {J.}~\bibnamefont {Kroha}},\ }\href
  {\doibase 10.1088/1464-4258/11/11/114012} {\bibfield  {journal} {\bibinfo
  {journal} {J. Opt. A: Pure Appl. Opt.}\ }\textbf {\bibinfo {volume} {11}},\
  \bibinfo {pages} {114012} (\bibinfo {year} {2009})}\BibitemShut {NoStop}%
\end{thebibliography}
%

\end{document}